\documentclass[12pt]{iopart}

\usepackage{iopams}
\usepackage{graphicx}
\usepackage{xcolor}
\usepackage{caption}
\usepackage{subcaption}
\usepackage[colorlinks=true,linkcolor=blue,citecolor=blue]{hyperref}
\usepackage{etoolbox}
\usepackage{cancel}
\usepackage{cite}
\usepackage{soul}

\patchcmd{\numparts}{\addtocounter{equation}{1}}{\refstepcounter{equation}}{}{}

\begin{document}

\title{Self-diffusion anomalies of an odd tracer in soft-core media}

\author{Pietro Luigi Muzzeddu$^{1,*}$, Erik Kalz$^{2,*}$, Andrea Gambassi$^{3,4}$, 
Abhinav Sharma$^{5,6, \dagger}$ and Ralf Metzler$^{2,7, \dagger}$}
\address{$^1$ University of Geneva, Department of Biochemistry, CH-1211 Geneva, 
Switzerland}
\address{$^2$ University of Potsdam, Institute of Physics and Astronomy, D-14476 
Potsdam, Germany}
\address{$^3$ SISSA - International School for Advanced Studies, IT-34136 Trieste, 
Italy}
\address{$^4$ INFN - Sezione di Trieste, IT-34127 Trieste, Italy}
\address{$^5$ University of Augsburg, Institute of Physics, D-86159 Augsburg, 
Germany}
\address{$^6$ Leibniz-Institute for Polymer Research, Institute Theory of Polymers, 
D-01069 Dresden, Germany}
\address{$^7$ Asia Pacific Centre for Theoretical Physics, KR-37673 Pohang, 
Republic of Korea}

\vspace{10pt}

\address{$^*$ These two authors contributed equally.}
\address{$^\dagger$ Corresponding authors: abhinav.sharma@uni-a.de, 
rmetzler@uni-potsdam.de}

\vspace{10pt}
\begin{indented}
\item[]\today
\end{indented}

\begin{abstract}
Odd-diffusive systems, characterised by broken time-reversal and/or parity
symmetry, have recently been shown to display counterintuitive features such
as interaction-enhanced dynamics in the dilute limit.  Here we we extend
the investigation to the high-density limit of an odd tracer embedded in a
soft-Gaussian core medium (GCM) using a field-theoretic approach based on the
Dean-Kawasaki equation.  Our theory reveals that interactions can enhance
the dynamics of an odd tracer even in dense systems.  We demonstrate that
oddness results in a complete reversal of the well-known self-diffusion
($D_\mathrm{s}$) anomaly of the GCM. Ordinarily, $D_\mathrm{s}$ exhibits
a non-monotonic trend with increasing density, approaching but remaining
below the interaction-free diffusion, $D_0$, ($D_\mathrm{s} < D_0$) so
that $D_\mathrm{s} \uparrow D_0$ at high densities. In contrast, for an odd
tracer, self-diffusion is enhanced ($D_\mathrm{s}> D_0$) and the GCM anomaly
is inverted, displaying $D_\mathrm{s} \downarrow D_0$ at high densities. The
transition between the standard and reversed GCM anomaly is governed by the
tracer's oddness, with a critical oddness value at which the tracer diffuses
as a free particle ($D_\mathrm{s} \approx D_0$) across all densities. We
validate our theoretical predictions with Brownian dynamics simulations,
finding strong agreement between the two.
\end{abstract}

\vspace{2pc}

\section{Introduction}
\label{sec:intro}

The transport of a tracer particle in crowded environments, such as the cytoplasm 
\cite{hofling2013anomalous} or the plasma membrane of cells 
\cite{ramadurai2009lateral} is of fundamental 
interest in biological applications \cite{scott2023extracting}.
This transport can be characterised by the 
mean-squared displacement (MSD) of the tracer particle 
which provides the self-diffusion coefficient for those systems that admit a 
long-time diffusive regime.
Self-diffusion as a concept 
goes back to the pioneering works by Maxwell \cite{maxwell1867iv}, 
Stefan \cite{stefan1871ueber} and Boltzmann \cite{boltzmann1896vorlesungen}. 
Already in 1875 Maxwell stated that \textit{``it is true that 
the diffusion of 
molecules goes on faster in a rarefied gas''} \cite{clerk1875dynamical}. 
This is not only intuitively appealing but has also been extensively 
validated in theoretical studies on
various interacting systems such as 
repulsive hard-spheres \cite{hanna1982self, imhof1995long, felderhof1983cluster}, 
Yukawa-like particles \cite{lowen1993long}, attractive Lennard-Jones-like 
colloids \cite{kushick1973role, bembenek2000role, yamaguchi2001mode, 
seefeldt2003self}, but also for bounded soft interactions \cite{wensink2008long, 
krekelberg2009anomalous}. Furthermore experimental studies have 
well-established the decrease of self-diffusion with interactions 
\cite{weeks2002properties, michailidou2009dynamics,peng2009short, 
thorneywork2015effect}.

Despite this seemingly paradigmatic effect, it is surprising that even the 
effect of repulsive interactions can enhance the self-diffusion 
\cite{kalz2022collisions}. This counter-intuitive behaviour is found in 
\textit{odd-diffusive} systems \cite{hargus2021odd, kalz2022collisions},  
characterised by probability fluxes perpendicular to the density gradient. The 
microscopic cause of the enhanced self-diffusion was attributed to 
a mutual rolling of particles induced by odd diffusion instead of the ordinary 
repulsion after an interaction \cite{kalz2022collisions, langer2024dance},
and associated with the non-Hermitian time evolution of the system 
\cite{kalz2024oscillatory}. Examples of odd systems include, for example, 
charged Brownian particles moving in the presence of a magnetic field and 
therefore subject to Lorentz force \cite{kurcsunoglu1962brownian, 
karmeshu1974brownian, chun2018emergence, park2021thermodynamic} at equilibrium 
but also active chiral particles \cite{muzzeddu2022active, chan2024chiral, 
kalz2024field, valecha2024chirality} out of equilibrium, diffusion in porous 
media structures \cite{koch1987symmetry, auriault2010asymmetry}, systems with 
transverse forces \cite{ghimenti2024irreversible, ghimenti2023sampling}, 
skyrmionic systems \cite{brown2018effect, gruber2023300, schick2024two}, 
optical tweezer experiments \cite{wu2009direct, volpe2023roadmap_short}, 
Magnus forces in soft matter \cite{reichhardt2022active, cao2023memory} and 
even the interstellar medium \cite{shalchi2011applicability, 
effenberger2012generalized} (see also the discussion in 
Ref.~\cite{langer2024dance}).

The phenomenon of interaction induced enhancement has been investigated 
only in the dilute limit and in combination with hard sphere interactions. 
Although molecular interactions always manifest a diverging repulsion at 
short distances, \emph{effective} soft interactions are known to emerge in 
numerous biophysical and soft matter systems~\cite{likos2001effective, 
vlassopoulos2014tunable}. A natural question now concerns the generality of the 
enhancement effect: does it persist at high densities and/or for a generic 
interaction potential? Both these questions can be addressed in the framework 
of a model with soft interactions where at high densities particles can 
interpenetrate. A very popular soft interaction potential is that of the Gaussian 
Core Model (GCM).
Pioneered by Stillinger in 1976 \cite{stillinger1976phase} and Stillinger and Weber 
\cite{stillinger1978study, stillinger1979high, stillinger1979duality,
stillinger1980lindemann, weber1981gaussian} starting from 1978, the model 
found wide applications while being recognised to accurately 
describe the effective interaction between polymer coils~\cite{flory1950statistical, 
louis1999structure, lang2000fluid, likos2001criterion, bolhuis2001accurate} over 
a wide range of densities~\cite{Louis2000can}, polymer-colloid mixtures
~\cite{bolhuis2002derive}, and flexible 
dendrimers~\cite{likos2001soft}. Interestingly, the GCM displays some 
counter-intuitive features which are referred to as static and dynamic 
anomalies~\cite{stillinger1997negative, mausbach2006static,mausbach2009transport, 
shall2010structural, 
krekelberg2009anomalous, speranza2011thermodynamic, archer2003statistical}, and 
its phase behaviour in the density-temperature plane has been exhaustively 
investigated in the last few 
decades~\cite{lang2000fluid, stillinger1997negative, sposini2023glassy, 
prestipino2011hexatic, 
mendoza2024melting}.
The GCM, for example, has an upper-freezing temperature 
above which the system is always fluid. There it has been shown that the 
diffusivity can increase upon 
isothermal compression until the GCM behaves as a ``high-density ideal gas'' 
\cite{lang2000fluid, krekelberg2009anomalous}.

In this work, we study via analytical and numerical methods the transport 
of an odd-diffusive tracer immersed in a medium of particles represented by 
the GCM.
To model the interactions of the tracer with the medium we employ a 
field-theoretic description within 
the Dean-Kawasaki approach~\cite{dean1996langevin, kawasaki1994stochastic, illien2024dean},
which allows us to make accurate analytical predictions specifically in highly 
dense, 
yet fluid systems.  
By comparing these predictions with 
Brownian dynamics simulations, we find remarkable agreement for two 
distinct 
regimes: depending on the strength of the oddness parameter $\kappa$, 
the interactions can either 
reduce the self-diffusion or enhance it. These regimes are 
separated by the oddness effect, where the host medium effectively is 
invisible to 
the tracer. Whereas dynamics can be enhanced in driven 
systems~\cite{benois2023enhanced, jardat2022diffusion, marbach2018transport, 
wang2023interactions}, it is remarkable that even in equilibrium systems such 
as those considered here, tracer diffusion can be enhanced by interparticle 
interactions. We further recover the diffusion anomaly for the GCM and find 
that in an 
odd-diffusive system, this anomaly as a function of the medium density is 
inverted when enhancement occurs.

The remainder of this work is organised as follows: in Section~\ref{sec:model} 
we set up the model and derive the governing time-evolution equations for the 
tracer and the host field. In Section~\ref{sec:self_diffusion} we employ a 
weak-coupling 
approximation and evaluate the self-diffusion within this perturbative approach. 
We the present the theoretical predictions and the results of 
numerical simulations. In Section~\ref{sec:conclusions} 
we present our conclusions and give an extended outlook to further 
applications of 
the effect. Additional details of the paper are presented in the appendices. 
In particular, in \ref{app:A} 
we discuss some subtleties in deriving the time-evolution of the 
odd tracer particle. In \ref{app:B} 
we solve the case in which the tracer does not interact with the medium. 
In \ref{app:C} we determine the first non-trivial 
contribution of the tracer-medium interaction to the self-diffusion within 
the weak-coupling 
approximation. \ref{app:simulations} provides details of the simulations. 

\begin{figure}[t]
    \centering
    \includegraphics[width=\textwidth]{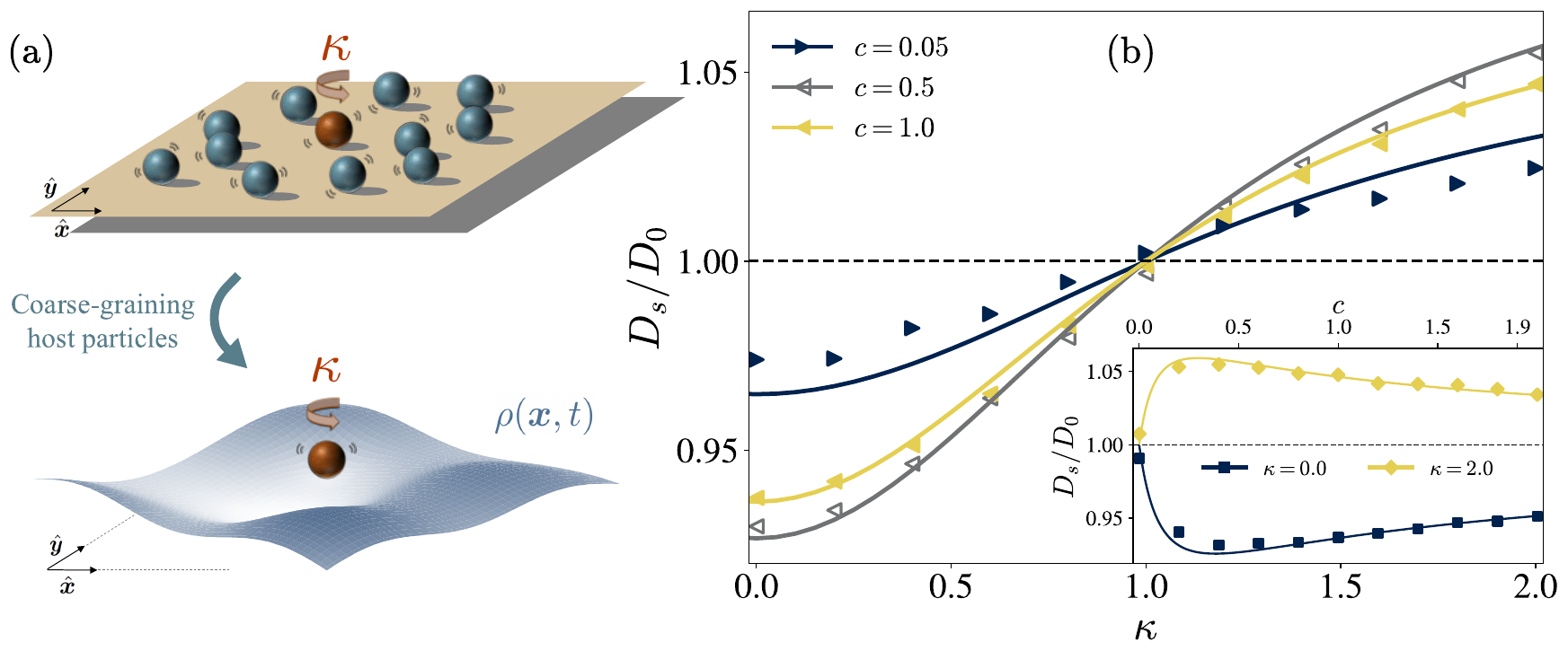}
    \caption{(a) Sketch of an odd-diffusing colloidal particle (red tracer 
    with $\kappa\neq 0$) coupled via soft-core interactions to a medium of normally 
    diffusing colloidal 
    particles (blue host particles with $\kappa = 0$). 
    In the mesoscopic description discussed here, the host particles are 
    accounted for in terms of the fluctuating density field $\rho(\bi{x},t)$. 
    (b) Analytical (perturbative) predictions (solid lines) are 
    compared with the results of 
    Brownian dynamics simulation (symbols) for the long-time self-diffusion 
    coefficient $D_{\mathrm{s}}/D_0$ of the odd tracer coupled to normal host 
    particles as a function of the oddness parameter $\kappa$, see 
    Eq.~\eref{eq:main_result_Ds_adimensional}. 
    For sufficiently small values of $\kappa$, the self-diffusion
    $D_\mathrm{s}$ is reduced compared to 
    the the value $D_0$ it takes for an odd tracer in the absence of 
    interactions with the host particles, as intuitively expected in a 
    crowded environment. For $\kappa>\kappa_c=1$, however, the interaction of 
    the tracer with the host particles enhances its self-diffusion. This 
    phenomenon persists for a remarkably wide range of the density $c$. Note that at 
    higher densities ($c=1.0$) the relative reduction/enhancement of the self 
    diffusion is 
    smaller than at intermediate densities ($c=0.5$). This can be understood by 
    the self-diffusion anomaly of the GCM (non-monotonic dependence of the 
    self-diffusion coefficient $D_\mathrm{s}$ as a function of the density, 
    highlighted in the inset of panel (b). The simulation parameters to obtain
     panel (b) are: $N=200$ particles of mass $m=0.01$, friction coefficient
    $\gamma=1.0$, thermal energy $T=1.0$, particle diametre $\sigma=1.0$, and
     $\bar{\lambda}=1.0$ (see definition in Eq.~\eref{eq:lambdabar}). For 
     further details on the simulations see~\ref{app:simulations}.}
    \label{fig:sketch_and_main_result}
\end{figure}

\section{The model}
\label{sec:model} 

We consider the stochastic dynamics in $d=2$ spatial dimensions of an 
ensemble of $N+1$ interacting particles 
with positions $\bi{X}_j(t) \in \mathbb{R}^2$ at time $t$, and 
$j \in \{ 0, \ldots, N\}$. A sketch of the system is shown in 
Fig.~\ref{fig:sketch_and_main_result}(a). 
The particle labelled by $\bi{X}_0$ is characterised by an 
odd-diffusive behaviour and its dynamics follow an underdamped equation of motion. We 
refer to this particle as the \textit{odd tracer}. In particular, the effect 
of oddness is encoded in the fact that its friction tensor 
$\bi{\Gamma}=\gamma [\mathbf{1} - 
\kappa \boldsymbol{\epsilon}]$ is characterised by antisymmetric elements 
proportional to the dimensionless oddness parameter $\kappa$, which stem from a 
non-conservative force experienced by the tracer \cite{chun2018emergence,
vuijk2019anomalous}. Here, $\gamma$ denotes the scalar friction coefficient. In 
the overdamped limit, this effect results in the odd diffusion tensor $\bi{D} 
= T \bi{\Gamma}^{-1} = D_0 (\mathbf{1} + \kappa\boldsymbol{\epsilon})$, where $D_0 
= T/(\gamma(1 + \kappa^2))$ is the scalar diffusion coefficient of a free odd particle 
\cite{chun2018emergence}, and 
$T$ is the temperature of the thermal bath measured in units such that the Boltzmann constant is 
one. We denote with $\boldsymbol{\epsilon}$ the two-dimensional antisymmetric Levi-Civita symbol 
($\epsilon_{xx}=\epsilon_{yy}=0$ and $\epsilon_{xy}=-\epsilon_{yx} = 1$) and with
$\mathbf{1}$ the identity matrix. The stochastic dynamics of the odd tracer can be 
written as
\begin{eqnarray}
    \label{eq:underdamped_dynamics_odd_tracer_position}
    \dot{\bi{X}}_0&=\bi{V}_0,\\
     m\dot{\bi{V}}_0&=-\lambda_\mathrm{tr} \sum_{j=1}^N \nabla \mathcal{U}
    (\bi{X}_0-\bi{X}_j) - \bi{\Gamma} \bi{V}_0 + \sqrt{2 T \gamma}\, 
    \boldsymbol{\xi}_0,
    \label{eq:underdamped_dynamics_odd_tracer}
\end{eqnarray}
where $\bi{V}_0(t)$ denotes the velocity of the odd tracer at time $t$, $m$ its 
mass, $\lambda_\mathrm{tr}$ is the overall strength of its interaction 
$\mathcal{U}$ with
the host particles, while $\boldsymbol{\xi}_0$ belongs to a set of $N+1$ independent 
zero-mean Gaussian white noises $\{ \boldsymbol{\xi}_i \}_{i=0}^N$ with correlation
\begin{equation}
        \langle \boldsymbol{\xi}_i(t) \otimes \boldsymbol{\xi}_j(s) \rangle= 
        \delta_{ij}\, \delta(t-s)\, \mathbf{1}.
    \label{eq:correlation_xi0}
\end{equation}
Here, the symbol $ \otimes $ denotes the outer product 
between two vectors $\bi{a}$ and $\bi{b}$, so that $[\bi{a} \otimes 
\bi{b}]_{\alpha \beta}=a_\alpha b_\beta$. 
The interaction potential $\mathcal{U}$ in 
Eq.~\eref{eq:underdamped_dynamics_odd_tracer} is assumed to be pairwise 
and with a smooth behaviour at the origin, such that $\nabla 
\mathcal{U}(\mathbf{0})=0$. Although we are ultimately interested in investigating 
the self-diffusion of the odd tracer in the overdamped regime, in which inertial 
effects can be neglected compared to viscous forces, we keep the time scale 
$\tau_\gamma=m/\gamma$  
finite throughout the derivation, and take the limit $m \to 0$ at the end of 
the calculation because the stochastic description of an overdamped odd particle 
bears some 
non-trivial subtleties \cite{chun2018emergence, park2021thermodynamic}. As 
shown in \ref{app:A}, the velocity $\bi{V}_0$ can be marginalised out at the 
level of the stochastic dynamics, leading to the following equation of motion 
for the position of the odd tracer
\begin{eqnarray}
\label{eq:colored_dynamics_position}
    \dot{\bi{X}}_0=&- \frac{\lambda_\mathrm{tr}}{m}\int_{t_0}^t \mathrm{d}s\, 
    \bi{G}(t-s) \sum_{j=1}^N\nabla \mathcal{U}(\bi{X}_0(s)-\bi{X}_j(s))  
    \nonumber \\ &+ \bi{G}(t-t_0) \bi{V}_0(t_0) +\boldsymbol{\eta}(t),
\end{eqnarray}
where the memory matrix $\bi{G}(u)$ is defined as
\begin{eqnarray}
    \label{eq:G_definition}
    \bi{G}(u)&= \mathrm{e}^{-|u|/\tau_\gamma} \bi{M}(u),\\
    \bi{M}(u)&= \left(\begin{array}{cc}
    \cos (u \kappa/\tau_\gamma) & \sin (u \kappa/\tau_\gamma) \\
    -\sin (u \kappa/\tau_\gamma) & \cos (u \kappa/\tau_\gamma)
    \end{array}\right)\,,
\label{eq:M_definition}
\end{eqnarray}
and $t_0$ is the time at which the initial conditions are imposed.
Odd diffusion thus introduces oscillations in the dynamics of the tracer
which decay on a time-scale 
$\tau_\gamma$ and vanish in the limit of a normal-diffusive system, i.e., 
$\bi{G}(u) \to \mathrm{exp}(-|u|/\tau_\gamma)\, \mathbf{1}$ as $\kappa 
\to 0$. The memory introduced by the coarse-graining of the velocity 
appears in the convolution of 
$\bi{G}(u)$ with the interaction forces, as well as in the zero-mean Gaussian 
coloured noise
\begin{equation}
    \boldsymbol{\eta}(t)=\frac{\sqrt{2T \gamma}}{m} \int_{t_0}^t \mathrm{d}s\, 
    \bi{G}(t-s) \boldsymbol{\xi}_0(s),
    \label{eq:colored_noise}
\end{equation}
with correlation 
\begin{equation}
    \langle \boldsymbol{\eta}(t) \otimes \boldsymbol{\eta} (s) \rangle=
    \frac{T}{m} \left[ \mathrm{e}^{-|t-s|/\tau_\gamma} - 
    \mathrm{e}^{-(t+s-2t_0)/\tau_\gamma} \right] \bi{M}(t-s).
    \label{eq:correlation_colored_noise}
\end{equation} 
At long times $t-t_0, s - t_0 \gg \tau_\gamma$, the effect of the initial 
conditions 
is forgotten and the 
two-point correlation function above becomes time-translation invariant and 
reads
\begin{eqnarray}
     \langle \boldsymbol{\eta}(t) \otimes \boldsymbol{\eta} (s) \rangle \simeq  
     \frac{T}{m} \bi{G}(t-s),\quad  t,s \gg t_0,
    \label{eq:correlation_colored_noise_TTinvariant}
\end{eqnarray}
which corresponds to the noise correlation reported in Ref. 
\cite{chun2018emergence}. 
In the same spirit as Refs.~\cite{jardat2022diffusion, 
benois2023enhanced}, we find it convenient 
to describe the particles constituting the host medium in terms of their 
density field.
Indeed, we are not interested in the microscopic details of the crowding host 
particles, but only in investigating the extent to which they affect the 
dynamics of the odd tracer. To 
this aim, following Ref. \cite{dean1996langevin}, we introduce the fluctuating 
particle density
\begin{equation}
    \rho(\bi{x},t)=\sum_{j=1}^N \delta(\bi{x}-\bi{X}_j(t))\,,
    \label{eq:fluctuating density}
\end{equation}
and use it in order to formulate a mesoscopic description of the medium based 
on the microscopic one.
The dynamics of the tracer can be specified from 
Eq.~\eref{eq:colored_dynamics_position} as
\begin{eqnarray}
\label{eq:dyanmics_tracer_field}
\dot{\bi{X}}_0&=- \frac{\lambda_\mathrm{tr}}{m}\int_{t_0}^t \mathrm{d}s\, 
\bi{G}(t-s)  \int \mathrm{d}\bi{x}\, \nabla \mathcal{U}(\bi{X}_0(s)-\bi{x})\, 
\rho(\bi{x},s) \nonumber \\& \quad + \bi{G}(t-t_0) \bi{V}_0(t_0) +
\boldsymbol{\eta}(t)\,.
\end{eqnarray}
The stochastic evolution of the density $\rho(\bi{x},t)$ appearing in 
the previous equation can be determined on the basis of
the microscopic dynamics of the $N$ host particles. In the overdamped 
regime the dynamics are
\begin{eqnarray}
    \dot{\bi{X}}_j&=-\frac{\lambda_\mathrm{ho}}{\gamma} \sum_{k=1}^N\nabla 
    \mathcal{U}(\bi{X}_j-\bi{X}_k) - \frac{\lambda_\mathrm{tr}}{\gamma}
    \nabla \mathcal{U}(\bi{X}_j-\bi{X}_0) 
    +\sqrt{2T/\gamma}\, 
    \boldsymbol{\xi}_j,
\end{eqnarray}
where $j \in \{1,...,N \}$. 
Here we assume that the potential which acts between the various pairs of 
host particles is, up to an overall constant 
$\lambda_\mathrm{ho}/\lambda_\mathrm{tr}$, the same as that which determines the 
interaction $\lambda_\mathrm{tr}\mathcal{U}$ between each host particle and the 
tracer.
Analogously to the original works by Kawasaki \cite{kawasaki1994stochastic} 
and Dean \cite{dean1996langevin}, the 
stochastic dynamics of $\rho(\bi{x},t)$ can be derived using It\^o's lemma, and 
it turns out to be governed by the following continuity equation
\begin{equation}
\frac{\partial}{\partial t} \rho(\bi{x},t)= - \nabla_{\bi{x}} \cdot 
\bi{\mathcal{J}}(\bi{x},t)\,,
\label{eq:density_dynamics}
\end{equation}
with the fluctuating flux 
\begin{eqnarray}
\bi{\mathcal{J}}(\bi{x},t)&= - \frac{\lambda_\mathrm{ho}}{\gamma} \int 
\mathrm{d}\bi{y}\, \rho(\bi{y},t)\, \nabla \mathcal{U}(\bi{x}-\bi{y})\, 
\rho(\bi{x},t) - \frac{\lambda_\mathrm{tr}}{\gamma}\, \nabla \mathcal{U}(\bi{x}-
\bi{X}_0)\, \rho(\bi{x},t) \nonumber\\
& \quad - \frac{T}{\gamma}\, \nabla \rho(\bi{x},t) + \sqrt{2T\, 
\rho(\bi{x},t)/\gamma}\, \boldsymbol{\Lambda}(\bi{x},t).
\label{eq:density_flux}
\end{eqnarray}
A few comments on the above equation are in order: the first line on the r.h.s. 
corresponds to the drift flux due to the soft interactions between the host 
particles in the medium, and to the interaction between the density field of the host 
particles and the odd tracer at position $\bi{X}_0$. 
The second line, instead, stems from the coupling of the density $\rho(\bi{x},t)$ 
with the equilibrium thermal bath at temperature $T$. This involves the standard 
diffusive flux proportional to $\nabla \rho $ and a fluctuating contribution 
that depends on the zero-mean Gaussian white noise field $\boldsymbol{\Lambda}
(\bi{x},t)$. The latter is characterised by the correlation
\begin{equation}
    \langle \boldsymbol{\Lambda}(\bi{x},t) \otimes \boldsymbol{\Lambda}(\bi{y},s) 
    \rangle = \delta(t-s)\, \delta(\bi{x}-\bi{y})\, \mathbf{1}. 
    \label{eq:correlation_noise_field}
\end{equation}
Note that the noise $\sqrt{2 T\, \rho(\bi{x},t)/\gamma}\,  
\boldsymbol{\Lambda}(\bi{x},t)$ is multiplicative as its amplitude depends on 
the fluctuating density itself and is to be interpreted in the It\^o sense. 
Moreover, Eq.~\eref{eq:density_dynamics} is 
nonlinear in the density $\rho$, thus it cannot be solved analytically.
In order to overcome this problem, we assume 
that the density fluctuations are much smaller than the 
homogeneous bulk density 
(see, e.g., Refs.~\cite{demery2014generalized, 
poncet2017universal, jardat2022diffusion, benois2023enhanced, venturelli2024universal}). 
In other words, 
we decompose the fluctuating density as $\rho(\bi{x},t)=\rho_0 + 
\phi(\bi{x},t)$, where $\rho_0 =N/L^2$ is the density in the 
homogeneous state, $N$ the number of particles and $L$ the typical box size,
while $\phi(\bi{x},t)$ represents the 
fluctuations around that state. Then we  assume that $|\phi(\bi{x},t)|\ll 
\rho_0$, 
which is increasingly accurate in the regime of high densities 
\cite{demery2014generalized}. At the lowest order in $\phi(\bi{x},t)/\rho_0$ 
from Eqs.~\eref{eq:density_dynamics} and \eref{eq:density_flux}, 
one gets the following linear dynamics for the field $\phi(\bi{x},t)$, 

\begin{eqnarray}
    \frac{\partial}{\partial t}\phi(\bi{x},t)&= \frac{\rho_0}{\gamma} \int 
    \mathrm{d}\bi{y}\, 
    \left[ \lambda_\mathrm{tr}\,  \delta(\bi{y}-\bi{X}_0) + \lambda_\mathrm{ho}\, 
    \phi(\bi{y},t)\right]\nabla_{\bi{x}}^2 \mathcal{U}(\bi{x}-\bi{y}) \nonumber \\
    & \quad +\frac{T}{\gamma}  \nabla^2 \phi(\bi{x},t) + \zeta(\bi{x},t),
    \label{eq:field_dynamics}
\end{eqnarray}
where we introduce the scalar zero-mean Gaussian white noise field 
$\zeta(\bi{x},t)$ with correlations 
\begin{equation}
    \langle \zeta(\bi{x},t) \zeta(\bi{y},s) \rangle = -\frac{2T\rho_0}{\gamma} 
    \delta(t-s)\,  \nabla_{\bi{x}}^2  \delta(\bi{x}-\bi{y}).
    \label{eq:correlation_zeta}
\end{equation}
Note that, in this context, $\rho_0/\gamma$ is often referred to as the 
field mobility coefficient~\cite{tauber2014critical}. In the regime of small 
density fluctuations, the microscopic equation of motion 
of the odd tracer derived from Eq.~\eref{eq:dyanmics_tracer_field} becomes
\begin{eqnarray}
\dot{\bi{X}}_0&=- \frac{\lambda_\mathrm{tr}}{m}\int_{t_0}^t \mathrm{d}s\, 
\bi{G}(t-s) \int \mathrm{d}\bi{x}\, \nabla \mathcal{U}(\bi{X}_0(s)-\bi{x})\, 
\phi(\bi{x},s) \nonumber\\
& \quad+ \bi{G}(t-t_0) \bi{V}_0(t_0) +\boldsymbol{\eta}(t)\,.
\label{eq:dyanmics_tracer_phi}
\end{eqnarray}

\section{Self-diffusion of the odd tracer}
\label{sec:self_diffusion}

\begin{figure}[t]
    \centering
    \includegraphics[width=\textwidth]{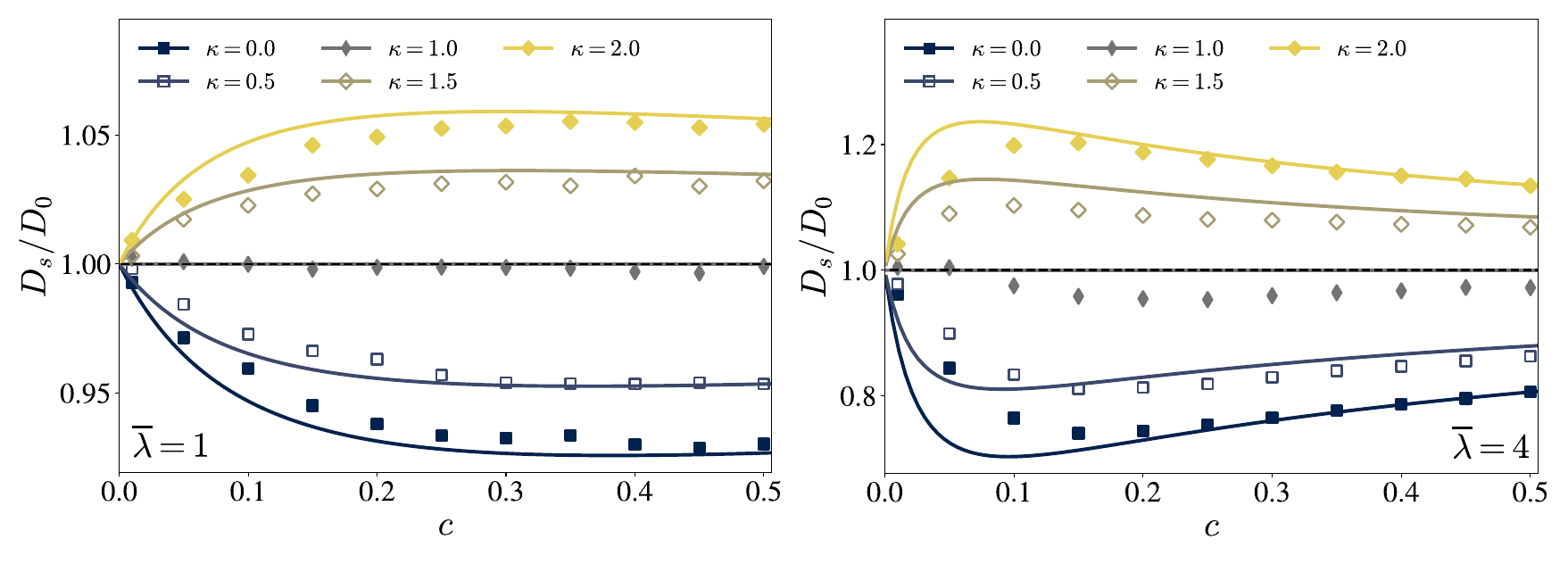}
    \caption{Analytical (perturbative) predictions (solid lines), see 
    Eq.~\eref{eq:main_result_Ds_adimensional}, and results of Brownian 
    dynamics simulation 
    results (symbols) for the long-time self-diffusion coefficient 
    $D_{\mathrm{s}}/D_0$ of an odd tracer coupled to normally diffusive host 
    particles as a function of the area fraction $c$ in two spatial 
    dimensions, with interparticle potential $\mathcal{U}$ given by 
    Eq.~\eref{eq:Gaussian_interaction_potential}. 
    (a) For an interaction coupling $\overline{\lambda}=1$ the 
    self-diffusion $D_\mathrm{s}$ decreases upon increasing density $c$ of 
    the medium for $\kappa<\kappa_c=1$, while it increases for 
    $\kappa >\kappa_c$. Interestingly, at the 
    critical value $\kappa_c$, the odd tracer diffuses at long times as in 
    the absence of host 
    particles, i.e., $D_\mathrm{s}=D_0$. (b) For a larger value of the 
    coupling $\overline{\lambda}=4$, we observe that the enhancement of the 
    self-diffusion 
    for $\kappa>\kappa_c$ is much more pronounced (up to $20\%$), and the 
    largest enhancement is obtained at a lower host medium density 
    ($c \simeq 0.15$). The self-diffusion anomaly of the GCM model 
    (i.e., the non-monotonic dependence of $D_\mathrm{s}$ on the host density) becomes 
    evident for $\kappa<\kappa_c$ and is inverted for $\kappa>\kappa_c$. 
    For the density getting higher, the GCM is known to exhibit a ``high-density 
    ideal gas''-like behaviour \cite{lang2000fluid}, a trend which we observe 
    here as well. The simulation parameters used to obtain the two panels are
    the same as in figure \ref{fig:sketch_and_main_result}.
    For further details on the simulations  see~\ref{app:simulations}.}
    \label{fig:Ds_vs_c}
\end{figure}

Using the evolution equations derived in the previous Section, we analyse the 
self-diffusion coefficient of the odd tracer and investigate how this is affected 
by the soft-core interactions with the host particles. 
Using a perturbative approach in the coupling strength $\lambda_\mathrm{tr}$ 
between the field $\phi(\bi{x},t)$ and the odd tracer, we compute the MSD of the 
latter and extract from its long-time behaviour the 
self-diffusivity defined as 
\begin{equation}
D_{\mathrm{s}}=\lim_{(t-t_0) \to \infty} \frac{\langle |\bi{X}_0(t)-\bi{X}_0
(t_0)|^2\rangle}{4(t-t_0)}\,.
\label{eq:def_self_diffusion}
\end{equation}
To this purpose, it is convenient to rewrite the stochastic dynamics of the 
field $\phi(\bi{x},t)$ in terms of its Fourier modes $\phi_{\bi{q}}(t)$, 
where the Fourier transform $f_{\bi{q}}$ of a function $f(\bi{x})$ is 
defined as $f_{\bi{q}} = \int\mathrm{d}
\bi{x}\ f(\bi{x})\, \mathrm{exp}(-i\bi{x}\cdot \bi{q})$. The field dynamics 
of Eq.~\eref{eq:field_dynamics} becomes
\begin{equation}
\frac{\partial}{\partial t}\phi_{\bi{q}}= -\alpha_{\bi{q}} \phi_{\bi{q}} 
-\lambda_\mathrm{tr}\, 
\frac{\rho_0}{\gamma} q^2 \mathcal{U}_{\bi{q}}\, \mathrm{e}^{-i\bi{q}\cdot 
\bi{X}_0} + \zeta_{\bi{q}}(t)\,,
\label{eq:dynamics_field_fourier}
\end{equation}
where we introduced the inverse relaxation time 
\begin{equation}
    \alpha_{\bi{q}}=
(\lambda_\mathrm{ho} \rho_0\, \mathcal{U}_{\bi{q}} +T)\, q^2/\gamma
\end{equation} 
of the
$\bi{q}$-mode of the field and the Fourier transform of the noise 
$\zeta_{\bi{q}}(t)$ with correlation
\begin{equation}
    \langle \zeta_{\bi{q}}(t)\zeta_{\bi{p}}(s) \rangle =  \frac{2 T\rho_0\, q^2}
    {\gamma} \delta(t-s)\, (2\pi)^2 \delta(\bi{q}+\bi{p})\,.
    \label{eq:correlation_phiq}
\end{equation}
Importantly, the relaxation time $1/\alpha_\mathbf{q}$ of $\phi_{\bi{q}}$ 
increases upon decreasing $q^2 \to 0$, 
eventually diverging for the $\bi{q}=\mathbf{0}$ mode. This is in agreement 
with the fact that the field $\phi(\bi{x},t)$ is a locally conserved quantity, 
which evolves according to the continuity equation given by 
Eq.~\eref{eq:field_dynamics}. Note that the field evolution in 
Eq.~\eref{eq:dynamics_field_fourier} is formally analogous to the one reported 
in Refs.~\cite{venturelli2022nonequilibrium, venturelli2022inducing, 
venturelli2023memory, venturelli2024stochastic} for the case of model B dynamics.
The coupling between the odd tracer and the field in 
Eq.~\eref{eq:dyanmics_tracer_phi} can also be rewritten in terms of the modes 
$\phi_{\bi{q}}(t)$ and becomes
\begin{eqnarray}
\dot{\bi{X}}_0(t)&=- \frac{\lambda_\mathrm{tr}}{m}\int_{t_0}^t \mathrm{d}s\, 
\bi{G}(t-s)  \int \frac{\mathrm{d}\bi{q}}{(2 \pi)^2}\, i\bi{q} 
\mathcal{U}_{\bi{q}}\, \phi_{\bi{q}}(s)\, \mathrm{e}^{i \bi{q}\cdot\bi{X}_0(s)} 
\nonumber\\
&\quad+ \bi{G}(t-t_0) \bi{V}_0(t_0) +\boldsymbol{\eta}(t)\,.
\label{eq:dynamics_tracer_phiq}
\end{eqnarray}
In the next Section, we use Eqs.~\eref{eq:dynamics_tracer_phiq} 
and~\eref{eq:dynamics_field_fourier} as the starting point for the 
weak-coupling approximation.

\subsection{Weak--coupling approximation}
\label{subsec:small_coupling}

To compute the MSD, we formally expand the tracer 
position and the density field in powers of the coupling strength 
$\lambda_\mathrm{tr}$.
This technique was put forward by Refs. \cite{basu2022dynamics,
venturelli2022inducing, venturelli2022nonequilibrium} and adopted from a 
perturbative expansion of the generating functional of the dynamics within a 
path-integral formalism \cite{demery2011perturbative, demery2014generalized,
demery2019driven, demery2023non}. The formal expansion of the tracer position 
and field reads
\begin{eqnarray}
    \label{eq:perturbative_expansion}
    \bi{X}_0(t) &= \sum_{n=0}^\infty \lambda_\mathrm{tr}^n \bi{X}_0^{(n)}(t)\,, \\
    \phi_{\bi{q}}(t)&=\sum_{n=0}^\infty \lambda_\mathrm{tr}^n 
    \phi_{\bi{q}}^{(n)}(t). \label{eq:perturbative_expansion_2}
\end{eqnarray}
Following Refs.~\cite{basu2022dynamics, venturelli2022inducing, 
venturelli2022nonequilibrium} we use Eqs.~\eref{eq:perturbative_expansion} 
and~\eref{eq:perturbative_expansion_2} to obtain a series 
expansion for the MSD of the tracer, 
\begin{eqnarray}
    \langle |\bi{X}_0(t)|^2\rangle & = \left\langle \bi{X}_0^{(0)}(t)\cdot 
    \bi{X}_0^{(0)}(t)\right\rangle \nonumber\\
    & \quad + \lambda_\mathrm{tr}^2 \left( \left\langle\bi{X}_0^{(1)}(t)\cdot 
    \bi{X}_0^{(1)}(t)\right\rangle +2\left\langle \bi{X}_0^{(0)}(t)\cdot 
    \bi{X}_0^{(2)}(t)\right\rangle \right)  \nonumber\\
    & \quad + \mathcal{O}(\lambda_\mathrm{tr}^4),
    \label{eq:MSD_correction}
\end{eqnarray}
up to corrections of order $\mathcal{O}(\lambda_\mathrm{tr}^4)$ and where 
we assumed that the initial position of the tracer is $\bi{X}_0(t_0)=\mathbf{0}$ 
without loss of generality. However, the accuracy of the results obtained 
with the truncated series has to be checked \textit{a posteriori} with 
numerical simulations. Importantly, all contributions related to odd 
powers of $\lambda_\mathrm{tr}$ vanish as the equations of motion 
are invariant under the transformation 
$(\lambda_\mathrm{tr},\phi) \leftrightarrow (-\lambda_\mathrm{tr},-\phi)$ 
\cite{basu2022dynamics, venturelli2022nonequilibrium}.
In order to evaluate the MSD we thus have to solve the set of coupled 
stochastic 
dynamics for the tracer and the field, Eqs.~\eref{eq:dynamics_tracer_phiq} 
and~\eref{eq:dynamics_field_fourier}, respectively, at different orders 
of the expansion in $\lambda_\mathrm{tr}$. At the lowest order 
$\mathcal{O}(\lambda_\mathrm{tr}^0)$ we find
\begin{equation}
    \dot{\bi{X}}^{(0)}_0(t) = \bi{G}(t-t_0) \bi{V}_0(t_0) +\boldsymbol{\eta}(t)\,,
    \label{x0_time_evolution_equation}
\end{equation}
which determines the time evolution of the tracer particle in the 
absence of the interaction of the medium and 
is solved in \ref{app:B}. The tracer evolution to linear order 
$\mathcal{O}(\lambda_\mathrm{tr}^1)$ becomes
\begin{eqnarray}
    \dot{\bi{X}}_0^{(1)}(t) &= - \frac{1}{m} \int_{t_0}^t \mathrm{d}s\ 
    \bi{G}(t-s) \int \frac{\mathrm{d}\bi{q}}{(2\pi)^2}\,  i\bi{q}\, 
    \mathcal{U}_{\bi{q}}\, \phi_{\bi{q}}^{(0)}(s)\,  \mathrm{e}^{i \bi{q}
    \cdot \bi{X}_0^{(0)}(s)},
    \label{eq:X_first_order}
\end{eqnarray}
which depends on the free field $\phi_{\bi{q}}^{(0)}$ and the free 
tracer $\bi{X}_0^{(0)}$. 
Similarly, at order $\mathcal{O}(\lambda_\mathrm{tr}^2)$ we find
\begin{eqnarray}
    \dot{\bi{X}}_0^{(2)}(t)= -\frac{1}{m} \int_{t_0}^t \mathrm{d}s\ 
    &\bi{G}(t-s) \int \frac{\mathrm{d}\bi{q}}{(2\pi)^2}\, i\bi{q}\, 
    \mathcal{U}_{\bi{q}} \\
    &\times\left[\phi_{\bi{q}}^{(1)}(s) + i\bi{q}\cdot\bi{X}_0^{(1)}(s)
     \,\phi_{\bi{q}}^{(0)}(s)\right] \mathrm{e}^{i \bi{q}\cdot \bi{X}_0^{(0)}(s)}
     \nonumber,
\end{eqnarray}
which, again, is related to the tracer position and the field at lower orders 
in the interaction coupling. The relevant correlations within the weak-coupling 
approximations, i.e., $\langle\bi{X}_0^{(1)}(t)\cdot \bi{X}_0^{(1)}(t)\rangle$ 
and $\langle \bi{X}_0^{(0)}(t)\cdot \bi{X}_0^{(2)}(t)\rangle$, are evaluated in 
\ref{app:C}, Eqs.~\eref{eq:computing_c1_2} and \eref{eq:computing_c2} 
respectively. In particular, in the overdamped regime $m \to 0$, the 
expressions for $\langle\bi{X}_0^{(1)}(t)\cdot \bi{X}_0^{(1)}(t)\rangle$ 
and $\langle \bi{X}_0^{(0)}(t)\cdot \bi{X}_0^{(2)}(t)\rangle$ simplify 
significantly, Eqs.~\eref{small_mass_limit_ofx1x1} and \eref{small_mass_limit_ofx0x2} 
respectively. There our theoretical approach yields 
\begin{equation}
    \frac{D_{\mathrm{s}}}{D_0}=1 - \frac{\lambda_\mathrm{tr}^2 \rho_0}{2 
    \gamma^2} \frac{1 - \kappa^2}{1+\kappa^2} \int  \frac{\mathrm{d} \bi{q}}
    {(2\pi)^2} \frac{q^4 |\mathcal{U}_{\bi{q}}|^2}{\alpha_{\bi{q}
    }(\alpha_{\bi{q}
    }+D_0q^2)}+\mathcal{O}(\lambda_\mathrm{tr}^4)\, 
    \label{eq:main_result_Ds}
\end{equation}
for the self-diffusion coefficient, which constitutes the central result of the 
present work. 

This equation predicts that the sign of the first non-trivial perturbative 
correction to the self-diffusion $D_{\mathrm{s}}$ is solely governed by the 
oddness parameter $\kappa$, due to the proportionality to the factor $(1-\kappa^2)$. 
This implies that the critical value $\kappa_c=1$ separates two distinct regimes 
in which the interaction with the host medium suppresses ($\kappa<\kappa_c$) or 
enhances ($\kappa>\kappa_c$) the self-diffusion of the odd tracer; in 
particular, for $\kappa=\kappa_c$ we observe $D_\mathrm{s}=D_0$. Note that the 
specific value $\kappa_c=1$ of the critical oddness parameter, as well as the 
counter-intuitive independence of $\kappa_c$ on the density of the host medium, 
is expected to be accurate within the weak-coupling regime only. Furthermore, 
it is apparent from Eq.~\eref{eq:main_result_Ds} that for some choices of the 
parameters the corrected self-diffusion $D_s$ becomes negative. This unphysical 
behaviour suggests that in those regions of the parameter space the weak-coupling 
approximation breaks down. Another important aspect to highlight 
is the fact that the derivation of Eq.~\eref{eq:main_result_Ds} does not rely on 
the specific shape of the interaction potential $\mathcal{U}$.

\subsection{Gaussian-core model}
\label{subsec:gaussian_core_model}

We proceed and specialise the predictions in Eq.~\eref{eq:main_result_Ds} 
for the case of the Gaussian core model (GCM). Note that even though the 
host-host and host-tracer interactions can be 
different in principle, we restrict our analysis to identical interactions and 
coupling strengths $\lambda_\mathrm{ho} = \lambda_\mathrm{tr} = \lambda$. In 
the GCM, particles interact via the (bounded) Gaussian interaction potential
\begin{equation}
    \mathcal{U}(\bi{x}) = \frac{1}{2\pi \sigma^2} \exp \left(
        -\frac{\bi{x}^2}{2 \sigma^2}\right).
    \label{eq:Gaussian_interaction_potential}
\end{equation}

The typical length scale of interaction is set by $\sigma$, 
which corresponds to the inter-particle 
distance at which the interaction force $\bi{F}(\bi{x}) = - \lambda\nabla  
\mathcal{U}(\bi{x})$ is the strongest.
Accordingly, we interpret $\sigma$ as an effective particle radius, which, in 
turn, defines a particle area of $\pi \sigma^2$ and thus an effective area 
fraction $c=\pi \sigma^2 \rho_0 =\pi \sigma^2 N/L^2$. In the framework of the 
mesoscopic field theoretic 
description introduced in Sec.~\ref{sec:model}, the length scale $\sigma$ also 
determines the range of interaction between the odd tracer and the fluctuating 
density field $\phi(\bi{x},t)$.  
We note that, as apparent from Eq.~\eref{eq:Gaussian_interaction_potential}, 
$\mathcal{U}$ has units $[\mathcal{U}]=1/\mathrm{m}^2$ and therefore $\lambda$ 
is dimensional with units $[\lambda]=\mathrm{J m^2}$. For it to become an 
expansion parameter it thus needs to be made dimensionless by a
typical energy and length scale of the system.
To this aim, we use the length scale $\sigma$ and the thermal energy, obtaining
\begin{equation}
\label{eq:lambdabar}
\lambda \to \overline{\lambda} = \frac{\lambda}{2\pi\sigma^2\, T},
\end{equation}
such that $[\overline{\lambda}]=1$, defining $T$ to be measured in units 
such that the Boltzmann constant is unity. In related works, 
the phase diagram of the GCM is analysed and $1/\overline{\lambda}$ is used as 
an effective temperature~\cite{stillinger1978study, likos2001criterion, lang2000fluid}. 
The GCM, thereby, was found to exhibit an upper-freezing 
temperature at $1/\overline{\lambda}_\mathrm{freez} = 0.0102$
~\cite{lang2000fluid}, above which it 
behaves as a fluid for all densities. Our parameter choice in 
Figs.~\ref{fig:sketch_and_main_result} and \ref{fig:Ds_vs_c} ensures that 
the medium is in the fluid phase with $\overline{\lambda} \ll 
\overline{\lambda}_\mathrm{freez}$, see also \ref{app:simulations}. 
The self-diffusion correction reported in Eq.~\eref{eq:main_result_Ds} can 
now be expressed in terms of dimensionless quantities as
\begin{equation}
    \frac{D_{\mathrm{s}}}{D_0}=1 - 2\pi\, \overline{\lambda}^{2} c\, 
    \frac{1 - \kappa^2}{1+\kappa^2} \int  \frac{\mathrm{d} \bi{p}}{(2\pi)^2} 
    \frac{ |\mathcal{U}_{\bi{p}/\sigma}|^2}{\beta_{\bi{p}/\sigma}(
        \beta_{\bi{p}/\sigma}+\frac{1}{1 + \kappa^2})}+\mathcal{O}(
            \overline{\lambda}^{4})\,,
    \label{eq:main_result_Ds_adimensional}
\end{equation}
where we introduced the rescaled wave vector $\bi{p}=\sigma \bi{q}$ and the 
dimensionless quantity $\beta_{\bi{q}}=(2\overline{\lambda} c \, 
\mathcal{U}_{\bi{q}} +1)$. Note that $\beta_{\bi{q}}$ is invariant under the 
transformation $(\overline{\lambda}\to a\overline{\lambda}, c \to c/a)$ 
where $a>0$ is a constant coefficient, i.e., it assumes the same 
value if we consider a more diluted medium with stronger interaction coupling. 
Consequently, it is easy to verify that the first non-trivial correction to the 
self-diffusion depends linearly on $a$. Remember further that 
$D_0 = T/(\gamma(1 + \kappa^2))$ which leaves the factor of $1/(1 + \kappa^2)$ 
in the integrand of Eq.~\eref{eq:main_result_Ds_adimensional}.

By specialising Eq.~\eref{eq:main_result_Ds_adimensional} to the GCM case and 
resorting to standard numerical integration schemes (see ~\ref{app:C}) for the 
evaluation of the momentum integral, 
we analyse the extent to which the long-time 
self-diffusion $D_\mathrm{s}$ of the odd tracer is affected by the 
interactions with the host medium. In Fig.~\ref{fig:sketch_and_main_result}(b) 
we show the behaviour of 
$D_{\mathrm{s}}$ as a function of the oddness parameter $\kappa$ and 
compare the theoretical predictions with Brownian dynamics simulations 
(see \ref{app:simulations} for details). 
In particular, we observe that in the weak-coupling limit, 
when the oddness parameter $\kappa$ is smaller than 
its critical value 
$\kappa_c=1$, for all densities $c$ the self-diffusion of the odd tracer 
is suppressed compared to the value $D_0$, 
which characterises 
a single odd particle in the absence of interactions. This behaviour is 
expected and can be observed in various systems with repulsive interactions, 
e.g., in systems with hard-core \cite{hanna1982self, imhof1995long, 
felderhof1983cluster, lowen1993long} or soft-core \cite{wensink2008long, 
krekelberg2009anomalous} repulsion, but also for attractive Lennard-Jones particles 
\cite{kushick1973role, bembenek2000role, yamaguchi2001mode, seefeldt2003self}. 
The decrease of the self-diffusion upon increasing the density of the 
particles in the medium reflects 
the intuition that in a crowded environment the motion of a diffusive tracer 
is hindered by the collisions with the other particles. 

However, the combined effect of the particle interactions with the 
odd-diffusive motion of the tracer 
eventually results in an inversion of this tendency. 
For $\kappa>\kappa_c$, we observe an enhancement of the self-diffusion 
$D_{\mathrm{s}}$ compared to $D_0$, irrespective of the area fraction $c$. 
This 
phenomenology was first observed for the self-diffusion of odd-diffusive hard 
disks in Ref. \cite{kalz2022collisions} and since then was confirmed via 
theory and 
simulations for systems of particles with excluded volume 
\cite{kalz2024oscillatory, 
langer2024dance, schick2024two}. 
Our finding proves that an enhancement of the self-diffusion is 
possible not only for systems of particles with hard-core repulsion, 
but also with generic bounded soft-core potentials, as shown in 
Eq.~\eref{eq:main_result_Ds}. 
At the same time, the fact of having soft-core interactions shifts the value 
of the critical oddness parameter to $\kappa_c=1$, while it was $\kappa_c=
1/\sqrt{3} \approx 0.58$ in the case of hard disks \cite{kalz2022collisions}. 
This implies that in the case of soft-core interactions 
a more pronounced chirality is required to enhance the transport properties 
of the odd-tracer, as the effect of the collisions is ``milder'' in the 
presence of soft interactions.

One of the main advantages of the field-theoretic description adopted here 
compared to the geometric approach of Ref.~\cite{kalz2022collisions}, is that 
it allows us to extend the investigation to the case of dense systems. In 
particular, we show 
in Fig.~\ref{fig:sketch_and_main_result}(b) the dependence of $D_\mathrm{s}$ 
on $\kappa$ for very dilute ($c=5\%$, 
dark blue), moderately dense ($c=50\%$, grey), and very dense ($c=100\%$, 
yellow) systems of host particles. 
It appears that the behaviour of the self-diffusion as a function of the density 
is far from trivial. Focusing 
on values of $\kappa$ with $\kappa<\kappa_c$, 
$D_{\mathrm{s}}$ is larger when the tracer is dispersed in a very dense medium 
($c=100\%$) than in the case of intermediate density ($c=50\%$). This seemingly 
counter-intuitive phenomenon of the GCM,
referred to as the ``self-diffusion 
anomaly'', is associated with the structural anomaly of the fluid GCM 
at high density~\cite{krekelberg2009anomalous, mausbach2006static, 
mausbach2009transport, 
shall2010structural, wensink2008long, pond2009composition, pond2011communication}.
Specifically, as the interaction potential is bounded, particles tend to 
overlap for sufficiently high densities, generating an entropic gain 
\cite{krekelberg2009anomalous}, and actually tend to display a 
``high-density ideal gas''-like 
behaviour for even higher densities \cite{lang2000fluid}, where $D_\mathrm{s} 
\uparrow D_0$.
In the inset of Fig.~\ref{fig:sketch_and_main_result}(b), we show the 
emergence of this anomaly by investigating the dependence of $D_{\mathrm{s}}$ 
on $c$ over a wide 
range of densities up to $c=200\%$. 
For a normal diffusive tracer ($\kappa=0$), we confirm the non-monotonic 
anomalous behaviour of $D_{\mathrm{s}}$, which first sharply decreases 
and then slowly increases upon increasing $c$. However, $D_\mathrm{s} 
< D_0$ for all densities $c$. 
Surprisingly, however, for $\kappa > \kappa_c =1$, we observe a specular trend 
showing 
an initial increase of the self-diffusion $D_\mathrm{s}$ for sufficiently dilute 
systems, 
followed by a decrease as a function of $c$, when its value exceeds a 
certain threshold. Thus, for $\kappa > \kappa_c$, the self-diffusion anomaly of 
the GCM is inverted, $D_\mathrm{s} \downarrow D_0$, and remarkably 
$D_\mathrm{s} > D_0$ for all densities.

In Fig.~\ref{fig:Ds_vs_c} we plot the self-diffusion $D_\mathrm{s}$ as a 
function of the 
density $c$  
up to $c=50\%$ for different values of the oddness parameter $\kappa$. In 
particular, panel 
(a) shows the predictions for a coupling parameter 
$\overline{\lambda}=1$ 
while panel (b) for $\overline{\lambda}=4$.
Notably, the analytical predictions (solid lines) are in excellent 
agreement with the results of Brownian dynamics simulations (symbols), 
especially for 
denser systems. This is coherent with the fact that the linearisation of 
the
Dean-Kawasaki equation in Eq.~\eref{eq:density_dynamics} 
provides more accurate results when the density fluctuations $\phi$ around 
the homogeneous bulk 
density $\rho_0$ are much smaller than $\rho_0$ itself. As anticipated by 
Fig.~\ref{fig:sketch_and_main_result}(b) and above, it turns out that 
$D_\mathrm{s} > D_0$ for $\kappa > \kappa_\mathrm{c}$, while $D_\mathrm{s} < D_0$ 
for $\kappa < \kappa_\mathrm{c}$. Particularly interesting 
in Fig.~\ref{fig:Ds_vs_c}(a) is the critical case $\kappa=\kappa_c$, for which the 
self-diffusion appears to be
insensitive to any variation of the medium density, and the tracer 
diffuses effectively as a free particle. This effect can be rationalised by 
noting that increasing the host density 
has a twofold effect on the tracer: on the one hand, it makes the surrounding 
environment more crowded, thus hindering the motion of the tracer; on the other 
hand, 
thanks to the mutual rolling mechanism analysed in Refs.~\cite{kalz2022collisions, 
langer2024dance}, the interaction with the host particles can speed up the 
dynamics of the odd tracer.
At $\kappa=\kappa_c$ these two 
effects balance each other and the tracer effectively evolves as in the absence 
of interactions. Also in this case, Fig.~\ref{fig:Ds_vs_c}(a) shows a 
remarkable agreement between analytical predictions and numerical 
data obtained from Brownian dynamics simulations. Importantly, the fact that 
$\kappa_c$ does not depend on the host medium density holds true only within 
the weak-coupling regime illustrated in panel (a) of Fig.~\ref{fig:Ds_vs_c}, 
while it is not longer the case when the interaction strength 
$\overline{\lambda}$ is increased, as shown in panel (b) of 
Fig.~\ref{fig:Ds_vs_c}.

With $\overline{\lambda}$ being chosen four times 
larger in Fig.~\ref{fig:Ds_vs_c}(b) than in Fig.~\ref{fig:Ds_vs_c}(a), 
the interaction-induced enhancement of 
$D_\mathrm{s}$ can reach $\approx 20\%$ of the free value $D_0$, compared to 
$\approx 5\%$ of Fig.~\ref{fig:Ds_vs_c}(a), which agrees with the 
$\overline{\lambda}$ rescaling of the interaction-correction observed in 
Eq.~\eref{eq:main_result_Ds_adimensional}. 
Upon increasing the value of the coupling $\overline{\lambda}$, the 
self-diffusion anomaly of the GCM is further observed starting from much 
lower densities ($c \approx 15\%$). In this case, the agreement between 
simulations and analytical predictions is slightly worse, specifically in 
the dilute regime due to the less accurate linearisation of Dean's equation. 
Further, the matching is less accurate in the dense regime when the oddness 
parameter approaches $\kappa_c$. This observation can be rationalised by 
noting that when the interaction coupling $\overline{\lambda}$ is 
sufficiently large and $\kappa \approx \kappa_c$, the first non-trivial 
correction to $D_{\mathrm{s}}$ in Eq.~\eref{eq:main_result_Ds_adimensional} 
becomes comparable with higher-order terms $\mathcal{O}(\overline{\lambda}^{4})$, 
which are neglected in our analytical treatment.

\section{Conclusions}
\label{sec:conclusions}
In this work, we studied the self-diffusion coefficient $D_\mathrm{s}$ of an 
odd-diffusive 
tracer coupled to an ensemble of normally diffusive crowding host particles. 
The pairwise interaction between the particles was modelled by the bounded soft-core 
Gaussian potential, the so-called Gaussian core model (GCM), implying that 
distinct particles may partially overlap. 
From the microscopic picture of the GCM fluid, we moved to a field-theoretic 
description based on the Dean-Kawasaki equation~\cite{dean1996langevin, 
kawasaki1994stochastic, illien2024dean} in which the host particles are coarse-grained into a 
thermally fluctuating density field $\rho(\bi{x},t)$. Under the assumption that 
the interaction coupling strength $\lambda$ between the density field and 
the odd tracer is sufficiently small, we obtained a perturbative 
expansion for the MSD of the latter, which we truncated at 
the first non-trivial order $\mathcal{O}(\lambda^2)$. Based on this expansion 
we evaluated 
the field-induced correction to the self-diffusion of the odd tracer and 
compared it with Brownian dynamics simulations. In particular, we showed that, 
upon increasing the oddness parameter $\kappa$, the collisions with the host 
particles have a substantially different effect on the transport properties of 
the tracer. Specifically, there exists a critical value $\kappa_c$ of the 
oddness parameter $\kappa$ such that 
for $\kappa<\kappa_c$ the self-diffusion $D_\mathrm{s}$ is reduced by the crowding 
effect introduced by the host particles, i.e., $D_\mathrm{s}$ decreases 
upon increasing the overall host particle density $c$. In contrast, for 
$\kappa>\kappa_c$ the 
interaction with the host particles leads to an enhancement of the self-diffusion 
upon increasing $c$. 
Moreover, we showed that this enhancement reaches its maximum at a 
specific density of 
the system, whose value depends on the interaction coupling $\lambda$. 
Beyond that value, $D_\mathrm{s}$ starts decreasing upon increasing $c$.
This non-monotonic behaviour of the self-diffusion as a function of the area 
fraction $c$ for $\kappa>\kappa_c$ is specular to that of normally diffusive 
Gaussian-core particles, for which the diffusion coefficient first 
sharply decreases and then slowly increases upon 
increasing the host density (see the self-diffusion anomaly of the GCM, 
discussed e.g., in 
Refs.~\cite{krekelberg2009anomalous, mausbach2006static, mausbach2009transport, 
shall2010structural, wensink2008long}). Finally, we showed that at the critical 
value $\kappa=\kappa_c$ and within the weak-coupling regime, the self-diffusion 
of the odd tracer is not affected 
by the collisions with the crowding particles, irrespective of their density.

The model presented here can be extended to address a variety of related problems. 
For example, the Dean-Kawasaki equation has already been generalised to 
the underdamped regime, by including in the description a momentum density 
field~\cite{nakamura2009derivation, duran2017general}. A potentially interesting 
direction is that of deriving the fluctuating hydrodynamic equations for a 
system of interacting odd-diffusive soft-core particles in the underdamped 
regime and then use it to study the dynamic behaviour of a tracer in such a 
medium. Moreover, 
the derivation presented here can be used for a systematic analysis of 
the role of mass on the transport properties of an odd tracer in a crowded 
environment, which we leave for future work.

The GCM is known to exhibit further anomalous properties beyond the 
self-diffusion anomaly discussed here, among which we mention 
a density anomaly (expansion upon isobaric cooling), a structural order 
anomaly (reduction of the short-range translational order upon isothermal 
compression) 
and a re-entrant melting transition for the solid GCM
~\cite{krekelberg2009anomalous, shall2010structural, mausbach2006static, 
speranza2011thermodynamic, lang2000fluid}. As we showed here, the interplay 
between the odd diffusivity of the tracer and the self-diffusion anomaly 
(increase upon isothermal compression) of 
the underlying fluid gives rise to an interesting and unexpected behaviour 
of the self-diffusion coefficient 
(see Fig.~\ref{fig:Ds_vs_c}(a)). It is therefore interesting to thoroughly 
investigate the influence of oddness on structural, transport and thermodynamic 
properties of the GCM. 

For a binary mixture of GCM particles, previous work showed the existence 
of fluid-fluid phase separation~\cite{archer2001binary,louis2000mean}. Coupling
odd tracers to this kind of mixtures might show surprising behaviour in 
the dynamics of the transition. 
correlations on arbitrarily large length scales which could introduce 
fluctuation-induced forces between the odd-particles~\cite{hertlein2008direct, 
gambassi2009critical, martinez2017energy, magazzu2019controlling, 
venturelli2022inducing, gambassi2024critical}. 
Finally, in the formulation presented here, the odd tracer and the host 
particles (and thus the density field $\rho$) evolve according to an equilibrium 
dynamics. It may be interesting to analyse the transport properties of an (odd) 
tracer coupled to an active (odd) fluid featuring nonequilibrium fluctuations, 
in which the detailed balance condition is not fulfilled. The odd medium 
could be 
described by continuum models based on fluctuating density and polarity fields, 
as for example in Ref.~\cite{kuroda2023microscopic}, or with hydrodynamic 
theories characterised by odd viscosity~\cite{fruchart2023odd}, where already 
remarkable effects for tracers have been 
reported \cite{reichhardt2019active, yang2021topologically, reichhardt2022active, 
poggioli2023odd, hosaka2023hydrodynamics, duclut2024probe,lier2023lift}.

\ack
P. L. M. acknowledges financial support from Erasmus+ for his 
3-month stay at the University of Potsdam from February til April 2024, P. L. M. 
and E. K. acknowledge financial support from Abhinav Sharma for their stay at 
the University of Augsburg in August 2023. P. L. M. and E. K. thank Davide 
Venturelli for insightful discussions on the Dean-Kawasaki equation. E. K., A. S., 
and R. M. 
acknowledge support by the Deutsche Forschungsgemeinschaft (grants No. SPP 2332 - 
492009952, SH 1275/5-1 and ME 1535/16-1).

\appendix

\section{Velocity marginalisation}
\label{app:A}

In this Appendix we derive the equation of motion for the position of the odd 
tracer reported in Eq.~\eref{eq:colored_dynamics_position} by integrating 
out the 
velocity variable $\bi{V}_0(t)$ from Eqs.~\eref{eq:underdamped_dynamics_odd_tracer_position} 
and \eref{eq:underdamped_dynamics_odd_tracer}. To this aim, we define the new 
variable $\bi{U}_0(t)=\bi{S}^{-1} \bi{V}_0(t)$, which is related to the velocity 
$\bi{V}_0(t)$ by a suitably chosen linear transformation $\bi{S}$. 
The latter has the 
property to diagonalise the friction tensor $\bi{\Gamma}$, and satisfies 
$\bi{S}^{-1}\, \bi{\Gamma} \bi{S}=\bi{L}$, with $\bi{L}, \bi{S} \in 
\mathbb{C}^{2 \times 2}$. In particular, $\bi{L}$ is diagonal and contains the 
eigenvalues of the friction tensor $L_{00}\equiv \ell_{0}=\gamma(1 - i \kappa)$ 
and $L_{11}\equiv \ell_{1}=\gamma(1 + i \kappa)$, where $i$ denotes the 
imaginary unit, while
\begin{equation}
    \bi{S}=\frac{1}{\sqrt{2}}
    \left(\begin{array}{cc}
        -i & i\\
        1 & 1
    \end{array}\right).
    \label{eq:varphi}
\end{equation}
Note that the emergence of complex eigenvalues is due to the oscillatory 
behaviour introduced by the oddness parameter $\kappa$. In the new variable 
$\bi{U}_{0}$, the 
dynamics of Eqs.~\eref{eq:underdamped_dynamics_odd_tracer_position} and 
\eref{eq:underdamped_dynamics_odd_tracer} can be formally solved, yielding
\begin{eqnarray}
\label{eq:formal_solution_u}
\bi{U}_{0}(t)& = \mathrm{e}^{-\frac{(t-t_0)}{m}\bi{L}}\, \bi{U}_{0}(t_0) -  
\frac{\lambda_\mathrm{tr}}{m} \int_{t_0}^t \mathrm{d}s\, \mathrm{e}^{-\frac{(t-s)}
{m}\bi{L}} \sum_{j=0}^N \bi{S}^{-1}\, \nabla \mathcal{U}(\bi{X}_0(s)-\bi{X}_j(s)) 
\nonumber\\
& \quad +  \frac{1}{m}\int_{t_0}^t \mathrm{d}s\, \mathrm{e}^{-\frac{(t-s)}{m}
\bi{L}}\,  \bi{S}^{-1}\, \boldsymbol{\xi}_{0}(s).
\end{eqnarray}
The expression for $\bi{U}_{0}(t)$ can be inverted back into the original 
variables to find the stochastic dynamics of the position $\dot{\bi{X}}_0(t)=
\bi{S}\bi{U}_0(t)$. Using the identity, obtained from straightforward 
computation,
\begin{equation}
    S_{\alpha \beta} S^{-1}_{\beta \sigma}\, f_\sigma(s)\,  \mathrm{exp}
    \left(-\frac{\ell_\beta}{m}(t-s)\right) = G_{\alpha \beta} (t-s)\,  f_\beta(s)
\end{equation}
where $\bi{f}(s)$ is a generic $2$-dimensional vector, $s<t$ and $\bi{G}(u)$ 
defined in Eq.~\eref{eq:G_definition} of the main text, 
Eq.~\eref{eq:colored_dynamics_position} is finally obtained. Note that 
repeated indices imply summation according to the Einstein notation. As 
described in 
Eq.~\eref{eq:colored_noise}, the evolution of the position of the odd tracer 
depends on the colored noise $\boldsymbol{\eta}(t)$, which is given by the 
convolution of the function $\bi{G}(u)$ with the white noise $\boldsymbol{\xi}_0$. 
The correlation of $\boldsymbol{\eta}(t)$ can be computed as 
\begin{eqnarray}
\label{eq:correlation_eta_1}
\langle \boldsymbol{\eta}(t)\otimes \boldsymbol{\eta}(s) \rangle &= 
\frac{2T \gamma}{m^2} \int_{t_0}^t \mathrm{d}t^\prime\int_{t_0}^s \mathrm{d}s'\, 
\bi{G}(t-t^\prime)\, \bi{G}^\mathrm{T}(s-s')\, \delta(t^\prime-s^\prime) 
\nonumber\\
& = \frac{2T \gamma}{m^2} \int_{t_0}^{\min {(t,s)}} \mathrm{d}t^\prime\, 
\bi{G}(t-t^\prime)\, \bi{G}^\mathrm{T}(s-t^\prime) \nonumber\\
& = \frac{T}{m} \left[ \mathrm{e}^{-|t-s|/\tau_\gamma} - 
\mathrm{e}^{-(t+s-2t_0)/\tau_\gamma} \right] \bi{M}(t-s),
\end{eqnarray}
which corresponds to the correlation reported in 
Eq.~\eref{eq:correlation_colored_noise} of the main text, and $\bi{M}$ is 
given by Eq. \eref{eq:M_definition}. For the matrix product 
in the above calculation we used the relation
\begin{equation}
    \bi{G}(t-t')\, \bi{G}^T(s-t')= \mathrm{e}^{-(t+s-2t^\prime)/\tau_\gamma}\, 
    \bi{M}(t-s),
    \label{eq:trigonometric_identity}
\end{equation}
which easily can be shown with the help of trigonometric identities.

\section{Tracer dynamics in the absence of interaction with the medium}
\label{app:B}

We here analyse the case in which the 
coupling between the odd tracer and the density field of the particles in 
the medium is switched off. We 
denote the position of such a free odd tracer as $\bi{X}_0^{(0)}$, and the free 
field as $\phi_{\bi{q}}^{(0)}$. 
\subsection{Free dynamics of the odd tracer} In the absence of the coupling to 
the field, i.e., $\lambda_\mathrm{tr}=0$, the stochastic dynamics of the 
odd tracer in Eq.~\eref{x0_time_evolution_equation} can be exactly solved, 
and yields
\begin{equation}
    \bi{X}^{(0)}_0(t)=\int_{t_0}^t \mathrm{d}s\, \bi{G}(s-t_0)\, \bi{V}_0(t_0) + 
    \int_{t_0}^t \mathrm{d}s\, \boldsymbol{\eta}(s),
    \label{eq:solution_free_tracer}
\end{equation}
where we assumed that the odd tracer is initially at 
$\bi{X}^{(0)}_0(t_0)=\mathbf{0}$, without loss of generality.  Note that 
$\bi{V}_0(t_0)$ is an assigned value and therefore does not need a 
perturbative 
expansion. As the position $\bi{X}^{(0)}_0$ of the tracer follows a Gaussian 
process, 
we can characterise it by its mean $\boldsymbol{\mu}_0(t)\equiv
\langle\bi{X}^{(0)}_0(t)\rangle$ and two-time connected correlation function 
$\bi{C}(t,s)$. From Eq.~\eref{eq:solution_free_tracer} these quantities can be 
straightforwardly obtained, yielding
\begin{equation}
\label{eq:mean_free_case}
\boldsymbol{\mu}_{0}(t)=m\, \bi{A}(t-t_0) \bi{V}_0(t_0),
\end{equation}
and
\begin{eqnarray}
\label{eq:correlation_free_case}
\bi{C}(t,s)&\equiv \left\langle \bi{X}_{0}^{(0)}(t) \otimes \bi{X}_{0}^{(0)}(s) 
\right\rangle - \left\langle \bi{X}_{0}^{(0)}(t)\right\rangle\, \otimes 
\left\langle \bi{X}_{0}^{(0)}(s)\right\rangle \nonumber\\
& =2 D_0\, \left[\min (s,t)-t_0\right]\, \mathbf{1}- m T 
\left[\bi{\Gamma}^{-1} \bi{A}(t-t_0) + (\bi{\Gamma}^{-1} \bi{A}(s-t_0))^
\mathrm{T}\right] \nonumber \\
& \quad +m T \left[ \Theta (t-s)\, \bi{\Gamma}^{-1} \bi{A}(t-s) + \Theta (s-t)\,  
(\bi{\Gamma}^{-1} \bi{A}(s-t))^\mathrm{T} \right] \nonumber \\
& \quad-mT \bi{A}(t-t_0) \bi{A}^\mathrm{T}(s-t_0),
\end{eqnarray}
where we introduced the abbreviation $\bi{A}(u)= \bi{\Gamma}^{-1} [\mathbf{1} - 
\bi{G}(u)]$. We denoted by $\bi{a} \otimes \bi{b} = a_\alpha b_\beta$ the outer 
product between two vectors $\bi{a}$ and $\bi{b}$, and we introduced the bare 
diffusion coefficient  $D_0 = T/(\gamma(1 + \kappa^2))$ of the odd tracer. 
Note that the connected 
correlation satisfies $\bi{C}(t,s)=\bi{C}^\mathrm{T}(s,t)$. Once 
$\boldsymbol{\mu}_{0}(t)$ and $\bi{C}(t,s)$ are known, we can compute the 
generating functional $\mathcal{Z}[\bi{j}]$ of the $n$-point correlations for 
the position of the odd tracer in the free case ($\lambda_\mathrm{tr}=0$),
\begin{equation}
    \mathcal{Z}[\bi{j}]=\left\langle \exp\left\{\int \mathrm{d}t\,\bi{j}(t)\cdot 
    \bi{X}^{(0)}_0(t) \right\}\right\rangle\,,
    \label{eq:generating_functional}
\end{equation}
where $\bi{j}(t)$ is an auxiliary field and the average is taken with respect 
to the Gaussian path probability
\begin{eqnarray}
\label{eq:path_probability_free case}
\mathcal{P}_0[\bi{x}] \propto \exp &\Bigg\{ -\frac{1}{2} \int \mathrm{d}t \int 
\mathrm{d}s\, [\bi{x}(t)-\boldsymbol{\mu}_0 (t)] \cdot \bi{C}(t,s)\, 
[\bi{x}(s)-\boldsymbol{\mu}_0(s)] \Bigg\}.
\end{eqnarray}
Solving the functional Gaussian integral in Eq.~\eref{eq:generating_functional} 
leads to the expression 
\begin{eqnarray}
\label{eq:generating_functional_2}
\mathcal{Z}[\bi{j}]=\exp &\Bigg\{\frac{1}{2} \int \mathrm{d}t \int \mathrm{d}s\, 
\bi{j}(t)\cdot  \bi{C}(t,s)\, \bi{j}(s) + \int \mathrm{d}t\, \bi{j}(t) \cdot 
\boldsymbol{\mu}_0(t) \Bigg\}
\end{eqnarray}
for the generating functional. The explicit expression of this generating 
functional will be particularly useful for deriving some of the expressions 
presented further below, see Eq.~\eref{eq:identity_2}.

\subsection{Free dynamics of the field}

In the absence of interactions with the tracer particle, the dynamics of the 
free field in Fourier space $\phi^{(0)}_{\bi{q}}$ follows an Ornstein-Uhlenbeck 
process, solved by
\begin{equation}
\label{eq:free_field_solution}
\phi_{\bi{q}}^{(0)}(t)=\phi_{\bi{q}}^{(0)}(t_0)\, \mathrm{e}^{-\alpha_{\bi{q}}
(t-t_0)}+\int_{t_0}^t \mathrm{d}s\,\mathrm{e}^{-\alpha_{\bi{q}}(t-s)}\, 
\zeta_{\bi{q}}(s).
\end{equation}
Here, $\phi_{\bi{q}}(t) = \int \mathrm{d}\bi{x}\ 
\phi(\bi{x},t)\ \mathrm{exp}(-i \bi{x}\cdot \bi{q})$ denotes the (two-dimensional) 
Fourier transform of a field $\phi(\bi{x},t)$ with wave 
vector $\bi{q}$.
From Eq.~\eref{eq:free_field_solution} we compute the two-point correlations 
of the free field as
\begin{eqnarray}
\label{eq:two_time_correlation_field}
&\left\langle \phi_{\bi{q}}^{(0)}(t)\, \phi_{\bi{p}}^{(0)}(s)  \right\rangle = 
\left\langle \phi_{\bi{q}}^{(0)}(t_0) \, \phi_{\bi{p}}^{(0)}(t_0) \right\rangle\, 
\mathrm{e}^{-\alpha_{\bi{q}}(t-t_0)}\, \mathrm{e}^{-\alpha_{\bi{p}}(s-t_0)} 
\nonumber \\
& \qquad \qquad+ \frac{T}{
    \lambda_\mathrm{ho} \mathcal{U}_{\bi{q}} + T/\rho_0} \left[ 
        \mathrm{e}^{-\alpha_{\bi{q}}|t-s|} - \mathrm{e}^{-\alpha_{\bi{q}}
        (t+s-2t_0)}  \right]\, (2\pi)^{2}\delta(\bi{q}+\bi{p}),
\end{eqnarray}
where we used that $\mathcal{U}_{\bi{q}} = 
\mathcal{U}_{-\bi{q}}$ which holds for any symmetric interaction potential. 
When $t=s$ and the field had sufficient time to relax (formally $t_0 \to -\infty$), 
the two-point correlator yields
\begin{equation}
\label{eq:stationary_variance_field}
    \left\langle \phi_{\bi{q}}^{(0)}(t)\, \phi_{\bi{p}}^{(0)}(t) \right\rangle=
    \frac{T}{
        \lambda_\mathrm{ho} \mathcal{U}_{\bi{q}} + T/\rho_0} \, (2\pi)^{2}
        \delta(\bi{q}+\bi{p}).
\end{equation}
Thus, if we assume that the field is distributed according to its equilibrium 
distribution before being put in contact with the odd tracer at time $t=t_0$, 
then $\langle \phi_{\bi{q}}^{(0)}(t_0)\, \phi_{\bi{p}}^{(0)}(t_0) 
\rangle$ is given by Eq.~\eref{eq:stationary_variance_field}. Under this 
assumption,  Eq.~\eref{eq:two_time_correlation_field} can be 
written as
\begin{eqnarray}
\label{eq:two_time_correlation_field_2}
\left\langle \phi_{\bi{q}}^{(0)}(t)\, \phi_{\bi{q}'}^{(0)}(s)  \right\rangle 
&=(2\pi)^{2}\delta(\bi{q}+\bi{p})\, \frac{T}{\lambda_\mathrm{ho} 
\mathcal{U}_{\bi{q}} + T/\rho_0} \mathrm{e}^{-\alpha_{\bi{q}}|t-s|} \nonumber\\
&\equiv(2\pi)^2 \delta(\bi{q}+\bi{p})\,  C_{\phi_{\bi{q}}}(t-s),
\end{eqnarray}
which defines the stationary time-translational invariant correlator 
$C_{\phi_{\bi{q}}}$ of the free field $\phi_{\bi{q}}^{(0)}$, which will be 
of importance later, see Eqs.~\eref{eq:identity_1} and \eref{eq:identity_2}.

\section{Weak-coupling approximation}
\label{app:C}

In this Appendix, we compute the first non-trivial perturbative correction to 
the MSD which, due to the symmetry $(\lambda_\mathrm{tr}, \phi) 
\leftrightarrow (-\lambda_\mathrm{tr}, -\phi)$ is of second 
order in the interaction coupling $\lambda_\mathrm{tr}$. In the case in which 
the tracer is initialised at the origin, i.e., for $\bi{X}_0(t_0)=\mathbf{0}$, 
this is 
formally given by Eq.~\eref{eq:MSD_correction}. In order to evaluate 
this correction, we need to separately compute the correlations 
$\langle\bi{X}_0^{(1)}(t)\cdot \bi{X}_0^{(1)}(t)\rangle$ and $\langle 
    \bi{X}_0^{(0)}(t)\cdot \bi{X}_0^{(2)}(t)\rangle$.
To evaluate the first, we formally solve the stochastic dynamics in 
Eq.~\eref{eq:X_first_order} to get
\begin{eqnarray}
\label{eq:computing_c1_1}
 &\left\langle\bi{X}_0^{(1)}(t)\cdot \bi{X}_0^{(1)}(t) \right\rangle = 
 -\frac{1}{m^2} \int \frac{d \bi{q}}{(2 \pi)^2} \int \frac{d \bi{p}}{(2 \pi)^2}\, 
 q_\beta p_\gamma\,\mathcal{U}_{\mathrm{t},\bi{q}}\, \mathcal{U}_{\mathrm{t},
 \bi{p}}
\nonumber\\
& \qquad\times \int_{t_0}^t \mathrm{d}s \int_{t_0}^s \mathrm{d}u \int_{t_0}^t 
\mathrm{d}s^\prime
\int_{t_0}^t \mathrm{d}u^\prime\,  G_{\alpha \beta} (s-u)\, G_{\alpha \gamma} 
(s^\prime-u^\prime)\nonumber \\
&\qquad\times \left\langle \phi_{\bi{q}}^{(0)}(u)\, \phi_{\bi{p}}^{(0)}
(u^\prime)\,  \mathrm{e}^{i \bi{q}\cdot \bi{X}_0^{(0)}(u)+i \bi{p}\cdot 
\bi{X}_0^{(0)}(u^\prime)}\right\rangle.
\end{eqnarray}
As the average in the last line only involves the position $\phi_{\bi{q}}^{(0)}$ 
of the free tracer 
$\bi{X}_0^{(0)}$ and the free field, it can be factorised 
as follows
\begin{eqnarray}
\label{eq:identity_1}
&\left\langle \phi_{\bi{q}}^{(0)}(u)\, \phi_{\bi{p}}^{(0)}(u^\prime)\, 
\mathrm{e}^{i \bi{q}\cdot \bi{X}_0^{(0)}(u)+i \bi{p}\cdot \bi{X}_0^{(0)}
(u^\prime)}\right\rangle \nonumber \\
& \qquad =\left\langle \phi_{\bi{q}}^{(0)}(u)\, \phi_{\bi{p}}^{(0)}(u^\prime) 
\right\rangle \left\langle \mathrm{e}^{i \bi{q}\cdot \bi{X}_0^{(0)}(u)+i 
\bi{p}\cdot \bi{X}_0^{(0)}(u^\prime)}\right\rangle \nonumber\\
& \qquad= (2 \pi)^2 \delta(\bi{q}+\bi{p})\,  C_{\phi_{\bi{q}}}(u-u^\prime)\,  
\mathcal{Q}_{\bi{q}}(u,u^\prime)
\end{eqnarray}
where we used Eq.~\eref{eq:two_time_correlation_field_2} and we introduced 
the two-time quantity $\mathcal{Q}_{\bi{q}}$, which can be obtained from the 
generating 
functional as 
\begin{eqnarray}
\label{eq:identity_2}
\mathcal{Q}_{\bi{q}}(u,u')&\equiv \mathcal{Z}[\bi{j}=i \bi{q} (\delta(t-u)-
\delta(t-u'))] \nonumber\\
&=\exp \Big\{ -\frac{1}{2}  \bi{q}\cdot[\bi{C}(u,u) + \bi{C}(u',u')]\bi{q} 
\nonumber\\ 
& \quad + \frac{1}{2} \bi{q}\cdot[ \bi{C}(u',u)+\bi{C}(u,u')]\bi{q} + i \bi{q} 
\cdot[\boldsymbol{\mu}_0(u)-\boldsymbol{\mu}_0(u')]\Big\}
\end{eqnarray}
Therefore, the correlation $\langle\bi{X}_0^{(1)}(t)\cdot \bi{X}_0^{(1)}(t) 
\rangle$ in Eq.~\eref{eq:computing_c1_1} can be rewritten as
\begin{eqnarray}
\label{eq:computing_c1_2}
&\left\langle\bi{X}_0^{(1)}(t)\cdot \bi{X}_0^{(1)}(t)\right\rangle = 
\frac{1}{m^2}\int \frac{d \bi{q}}{(2 \pi)^2}\, q_\beta q_\gamma\, 
|\mathcal{U}_{\mathrm{t},\bi{q}}|^2 \nonumber \\
& \qquad\times \int_{t_0}^t \mathrm{d}s \int_{t_0}^s \mathrm{d}u \int_{t_0}^t 
\mathrm{d}s^\prime
\int_{t_0}^{s^\prime} \mathrm{d}u^\prime\,  G_{\alpha \beta} (s-u)  
G_{\alpha \gamma}\,  (s^\prime-u^\prime) \nonumber\\
& \qquad \times  C_{\phi_{\bi{q}}}(u-u^\prime)\, \mathcal{Q}_{\bi{q}}(u,u^\prime).
\end{eqnarray}
Before calculating $\langle\bi{X}_0^{(0)}(t)\cdot \bi{X}_0^{(2)}(t) \rangle$ 
it is convenient to solve the dynamics of $\phi_{\bi{q}}^{(1)}$, obtaining
\begin{equation}
    \phi_{\bi{q}}^{(1)}(t)=-q^2 \mathcal{U}_{\bi{q}}\frac{\rho_0}{\gamma}
    \int_{t_0}^t\mathrm{d}s\,\mathrm{e}^{-\alpha_{\bi{q}}(t-s)}\, 
    \mathrm{e}^{-i\bi{q}\cdot \bi{X}^{(0)}_0(s)}\,,
    \label{eq:phi1_solution}
\end{equation}
where we used that $\phi^{(n)}_{\bi{q}}(t_0)=0$ for all $n \geq 1$. This is 
justified as we already assumed for Eq.~\eref{eq:stationary_variance_field} 
that the initial condition of the field is drawn from its equilibrium 
distribution in the absence of the coupling with the tracer. With the help 
of the 
identity (see Eq.~\eref{eq:generating_functional_2} for the definition of 
$\mathcal{Z}$)
\begin{eqnarray}
\label{eq:computing_c2}
&\left\langle X_0^{(0)}(t)\, \mathrm{e}^{-i \bi{q}\cdot [\bi{X}_0^{(0)}
(u^\prime)-\bi{X}_0^{(0)}(u)]}\right\rangle =\frac{\delta \mathcal{Z}[\{ \bi{j}\}]}
{\delta \bi{j}(t)} \Bigg|_{\bi{j}(t)=-i \bi{q} [\delta(t-u^\prime)-\delta(t-u)] } 
\nonumber\\
& \qquad =\mathcal{Q}_{\bi{q}}(u,u^\prime)\, \left[(\bi{C}(t,u)- 
\bi{C}(t,u^\prime))i\bi{q} + \boldsymbol{\mu}_0(t)\right],
\end{eqnarray}
the correlation $\langle\bi{X}_0^{(0)}(t)\cdot \bi{X}_0^{(2)}(t) \rangle$ can 
now be evaluated to be
\begin{eqnarray}
&\langle\bi{X}_0^{(0)}(t)\cdot \bi{X}_0^{(2)}(t)\rangle = \frac{i \rho_0}{\gamma m} 
\int\frac{\mathrm{d}\bi{q}}{(2\pi)^2} q_\beta q^2 |\mathcal{U}_{\mathrm{t},\bi{q}}|^2 
\nonumber\\ 
& \qquad \int_{t_0}^t \mathrm{d}s^\prime\int_{t_0}^{s^\prime} \mathrm{d}u^\prime 
\int_{t_0}^{u^\prime}dv^\prime\ G_{\alpha\beta}(s^\prime-u^\prime)\,  
\mathrm{e}^{-\alpha_{\bi{q}}(u^\prime-v^\prime)} \nonumber\\ 
&\qquad \times\left[\left(C_{\alpha\gamma}(t,u^\prime) - C_{\alpha\gamma}(t, 
v^\prime)\right) i q_\gamma  + \mu_0^\alpha(t)\right] \mathcal{Q}_{\bi{q}}
(u^\prime, v^\prime)\nonumber\\
&\quad + \frac{i}{m^2}\int\frac{\mathrm{d}\bi{q}}{(2\pi)^2} q^\beta q^\delta 
q^\epsilon |\mathcal{U}_{\mathrm{t},\bi{q}}|^2 \nonumber \\
&\qquad \int_{t_0}^t \mathrm{d}s^\prime \int_{t_0}^{s^\prime} \mathrm{d}u^\prime 
\int_{t_0}^{u^\prime} \mathrm{d}v^\prime \int_{t_0}^{v^\prime}\mathrm{d}w^\prime\ 
G_{\alpha\beta}(s^\prime-u^\prime)\, G_{\delta\epsilon}(v^\prime-w^\prime) 
\nonumber\\
&\qquad \times \left[\left(C_{\alpha\gamma}(t,u^\prime) - C_{\alpha\gamma}(t, 
w^\prime)\right)i q_\gamma + \mu_0^\alpha(t)\right]\, C_{\phi_{\bi{q}}}
(u^\prime-w^\prime) \, \mathcal{Q}_{\bi{q}}(u^\prime,w^\prime).
\label{eq:expression_c2}
\end{eqnarray}
The expression for the correlations given in Eqs.~\eref{eq:computing_c1_2} 
and~\eref{eq:expression_c2} are rather lengthy and do not admit an efficient 
numerical evaluation due to the nested time-integrals. However, these integrals 
can be analytically evaluated within the small-mass limit $m\to 0$ that 
characterises 
the overdamped regime. We can simplify the expression of $\mathcal{Q}_{\bi{q}}
(t,s)$ given in Eq.~\eref{eq:identity_2} by neglecting all contributions 
proportional to $m$ in its exponent and find
\begin{equation}
    \mathcal{Q}_{\bi{q}}(t,s) \to \mathrm{e}^{-q^2 D_0 |t-s| + i \bi{q}\cdot
    (\boldsymbol{\mu}_0(t)-\boldsymbol{\mu}_0(s))},\quad m\to 0,
    \label{eq:Q_small_mass}
\end{equation}
according to which $\mathcal{Q}_{\bi{q}}$ is an exponential function 
of the two times $t$ and $s$ only. Since also the two-point correlator 
$\bi{C}(t,s)$ of Eq.~\eref{eq:path_probability_free case}, the function 
$\bi{G}(t)$ defined in Eq.~\eref{eq:G_definition} of the main text, and the 
free-field correlator $C_{\phi_{\bi{q}}}(t)$ in 
Eq.~\eref{eq:two_time_correlation_field_2} can be rewritten as (complex) 
exponentials upon using suitable trigonometric identities, the nested 
time-integrals in Eqs.~\eref{eq:computing_c1_2} and~\eref{eq:expression_c2} can 
thus be solved analytically. Note that the validity of this seemingly 
uncontrolled approximation in Eq.~\eref{eq:Q_small_mass} is checked a 
posteriori by comparing the analytical 
predictions with numerical simulations. By further specialising this analysis 
to the long-time limit $t_0\rightarrow-\infty$, we rewrite the correlation 
in Eq.~\eref{eq:computing_c1_2} as
\begin{eqnarray}
&\left\langle\bi{X}_0^{(1)}(t)\cdot \bi{X}_0^{(1)}(t)\right\rangle =\frac{2 T 
 \rho_0}{\gamma m^2}(t-t_0) \int \frac{\mathrm{d} \bi{q}}{(2 \pi)^2}\, \frac{q^4 
|\mathcal{U}_{\mathrm{t},\bi{q}}|^2}{\alpha_{\bi{q}}} \mathrm{Re}[f_{\bi{q}}]\,,
    \label{eq:correlation_1_approx}
\end{eqnarray}
where $\mathrm{Re}[f_{\bi{q}}]$ denotes the real part of the momentum-dependent 
complex number $f_{\bi{q}}$ defined as 
\begin{eqnarray}
    f_{\bi{q}}=&\frac{\tau^2_\gamma}{\tilde{\alpha}_{\bi{q}}\left[1-\left( i 
    \kappa - \tau_\gamma \tilde{\alpha}_{\bi{q}}\right)^2 \right]} -
    \frac{\tau^4_\gamma\,  \tilde{\alpha}_{\bi{q}}}{ \left(1 + i\kappa \right)
    \left[\left(1 + i \kappa \right)^2 - (\tau_\gamma \tilde{\alpha}_{\bi{q}})^2 
    \right]}\,.
    \label{eq:fq}
\end{eqnarray}
To make the notation more compact, we further defined the new inverse time 
scale $\tilde{\alpha}_{\bi{q}}\equiv \alpha_{\bi{q}}+ D_0 q^2$ 
(compare with Eq.~\eref{eq:dynamics_field_fourier} for the definition of
 $\alpha_{\bi{q}}$). The correlation given in Eq.~\eref{eq:expression_c2} 
 can analogously be rewritten as
\begin{eqnarray}
\left\langle\bi{X}_0^{(0)}(t)\cdot \bi{X}_0^{(2)}(t)\right \rangle &= 
-\frac{2 D_0 \rho_0}{\gamma^2(1 + \kappa^2)}(t-t_0)\int \frac{\mathrm{d} \bi{q}}
{(2 \pi)^2}\,\frac{q^4 |\mathcal{U}_{\mathrm{t},\bi{q}}|^2}{\tilde{\alpha}^2_{\bi{q}}}
\nonumber\\
&\quad -\frac{2 \rho_0}{\gamma m}D_0^2(t-t_0) \int \frac{\mathrm{d} \bi{q}}
{(2 \pi)^2}\,\frac{q^6 |\mathcal{U}_{\mathrm{t},\bi{q}}|^2}{\alpha_{\bi{q}}} 
\mathrm{Re}[g_{\bi{q}}]\,,
    \label{eq:correlation_2_approx}
\end{eqnarray}
where we introduced the complex number $g_{\bi{q}}$ defined as
\begin{eqnarray}
    g_{\bi{q}}=\frac{\tau_\gamma\, (2\tau_\gamma\,  \tilde{\alpha}_{\bi{q}} 
    + 1-i \kappa)}{\tilde{\alpha}^2_{\bi{q}}\, \left( \tau_\gamma\, 
    \tilde{\alpha}_{\bi{q}} +1 -i \kappa \right)^2}\,.
    \label{eq:gq}
\end{eqnarray}
The remaining momentum integrals of Eqs.~\eref{eq:computing_c1_2} 
and~\eref{eq:expression_c2} can be finally 
performed numerically (e.g., using \texttt{Mathematica}). For the numerical 
evaluation, we truncated the integration domain $\mathbb{R}^2$ of the 
momentum integral into $[-q_b,q_b]^2$, by introducing the momentum cut-off 
$q_b=300$. The value of $q_b$, which guarantees an accurate estimate of 
the original integral, actually depends 
on the specific interaction potential, 
as well as on the other parameters of the model (see \ref{app:simulations} 
for more details). Here, we use a Gaussian interaction potential which displays 
a Gaussian decay on a momentum scale much smaller than $q_b$. We checked 
the validity of this approximation by testing the numerical integration for 
insensitivity against a variation of $q_b$ around the chosen value. By 
combining the numerical evaluation of Eqs.~\eref{eq:correlation_1_approx} 
and~\eref{eq:correlation_2_approx} with the formal expression of the first 
non-trivial 
perturbative correction to the MSD given in Eq.~\eref{eq:MSD_correction}, 
we obtain the results shown in panels (a) and (b) of Fig.~\ref{fig:Ds_vs_c}. 
Note that Eqs.~\eref{eq:correlation_1_approx} and~\eref{eq:correlation_2_approx} 
can be rewritten, in the overdamped regime $m\to0$, by using 
\begin{eqnarray}
    \frac{\mathrm{Re}[f_{\bi{q}}]}{m^2} \stackrel{m \to 0}{\longrightarrow} 
    \frac{1}{\gamma^2 \tilde{\alpha}_{\bi{q}}(1+\kappa^2)}\,,\\
    \frac{\mathrm{Re}[g_{\bi{q}}]}{m}\stackrel{m \to 0}{\longrightarrow} 
    \frac{1}{\gamma \tilde{\alpha}^2_{\bi{q}}(1+\kappa^2)}\,.
\end{eqnarray}
Specifically, these expressions allow us to rewrite 
Eq.~\eref{eq:correlation_1_approx} as
\begin{equation}
    \left\langle\bi{X}_0^{(1)}(t)\cdot \bi{X}_0^{(1)}(t)\right\rangle =\frac{2 D_0 
 \rho_0}{\gamma^2 }(t-t_0) \int \frac{\mathrm{d} \bi{q}}{(2 \pi)^2}\, \frac{q^4 
|\mathcal{U}_{\mathrm{t},\bi{q}}|^2}{\tilde{\alpha}_{\bi{q}}} \frac{1}{\alpha_{\bi{q}} }\,,
\label{small_mass_limit_ofx1x1}
\end{equation}
and Eq.~\eref{eq:correlation_2_approx} as
\begin{eqnarray}
\left\langle\bi{X}_0^{(0)}(t)\cdot \bi{X}_0^{(2)}(t)\right \rangle &= 
-\frac{2 D_0 \rho_0}{\gamma^2}(t-t_0)\int \frac{\mathrm{d} \bi{q}}
{(2 \pi)^2}\,\frac{q^4 |\mathcal{U}_{\mathrm{t},\bi{q}}|^2}{\tilde{\alpha}_{\bi{q}}}
\frac{1}{\tilde{\alpha}_{\bi{q}}(1 + \kappa^2)}
\\
&\quad -\frac{2 D_0\rho_0}{\gamma^2 }(t-t_0) \int \frac{\mathrm{d} \bi{q}}
{(2 \pi)^2}\,\frac{q^4 |\mathcal{U}_{\mathrm{t},\bi{q}}|^2}{
    \tilde{\alpha}_{\bi{q}}}\frac{D_0q^2}{\alpha_{\bi{q}} \tilde{\alpha}_{\bi{q}}
    (1+\kappa^2)} 
\,.\nonumber
\label{small_mass_limit_ofx0x2}
\end{eqnarray}
The sum of these two expressions can be easily computed and gives the result 
reported in Eq.~\eref{eq:main_result_Ds} of the main text.

\section{Simulations details}
\label{app:simulations}

\subsection{Brownian dynamics simulations.} 

The stochastic dynamics of the particles constituting the system can be 
conveniently cast in the general form of an underdamped Langevin equation for 
the $i$th particle,
\begin{eqnarray}
    \frac{\mathrm{d} \bi{X}_i(t)}{\mathrm{d}t} &= \bi{V}_i(t),\\
    \label{eq:velocity_SDE}
    m_i\frac{\mathrm{d} \bi{V}_i(t)}{\mathrm{d}t} &= -\gamma_i (\mathbf{1} - 
    \kappa_i \boldsymbol{\epsilon})\, \bi{V}_i(t) +  \bi{F}_i(t) +  
    \sqrt{2\gamma_i k_\mathrm{B}T}\, \boldsymbol{\xi}_i(t), 
\end{eqnarray}
where $\bi{X}_i(t), \bi{V}_i(t) \in \mathbb{R}^2$ are the $i$th particle 
position and velocity, 
respectively, with $i=0,1,\ldots N$ and a suitable choice of the parameters 
$m_i$, $\gamma_i$, $\kappa_i$. $\mathbf{1}$ is the identity matrix and 
$\boldsymbol{\epsilon}$ the two-dimensional Levi-Civita symbol. In total, we 
simulate the dynamics of $N=200$ particles, where the particle $i=0$ 
models the odd-diffusive tracer $(\kappa_0 =\kappa \neq 0$) and all other 
particles form the set of normal-diffusive 
host particles ($\kappa_i=0$ for $i = 1, \ldots, N$). The coefficients 
$\gamma_i$, $m_i$ denote the particles' friction and mass, respectively, and 
are assumed to be equal for the tracer and the host particles, i.e., 
$\gamma_i=1.0$ 
and $m_i=0.01$ for $i = 0, \ldots, N$. In units where the Boltzmann 
constant $k_\mathrm{B}$ is set 
to unity, the temperature of the thermal bath is taken to be $T=1.0$. 
$\boldsymbol{\xi}_i(t)$ is a zero-mean Gaussian white noise with correlations 
$\langle \xi_{i,\alpha}(t) \xi_{j,\beta}(t^\prime)\rangle = \delta_{ij}\, 
\delta_{\alpha\beta}\, \delta(t - t^\prime)$. Note that Latin indices $i,j$ 
label the particles, while Greek indices $\alpha, \beta$ refer to the 
(two) spatial 
coordinates. \par

\begin{figure}[t]
    \centering
   \includegraphics[width=\textwidth]{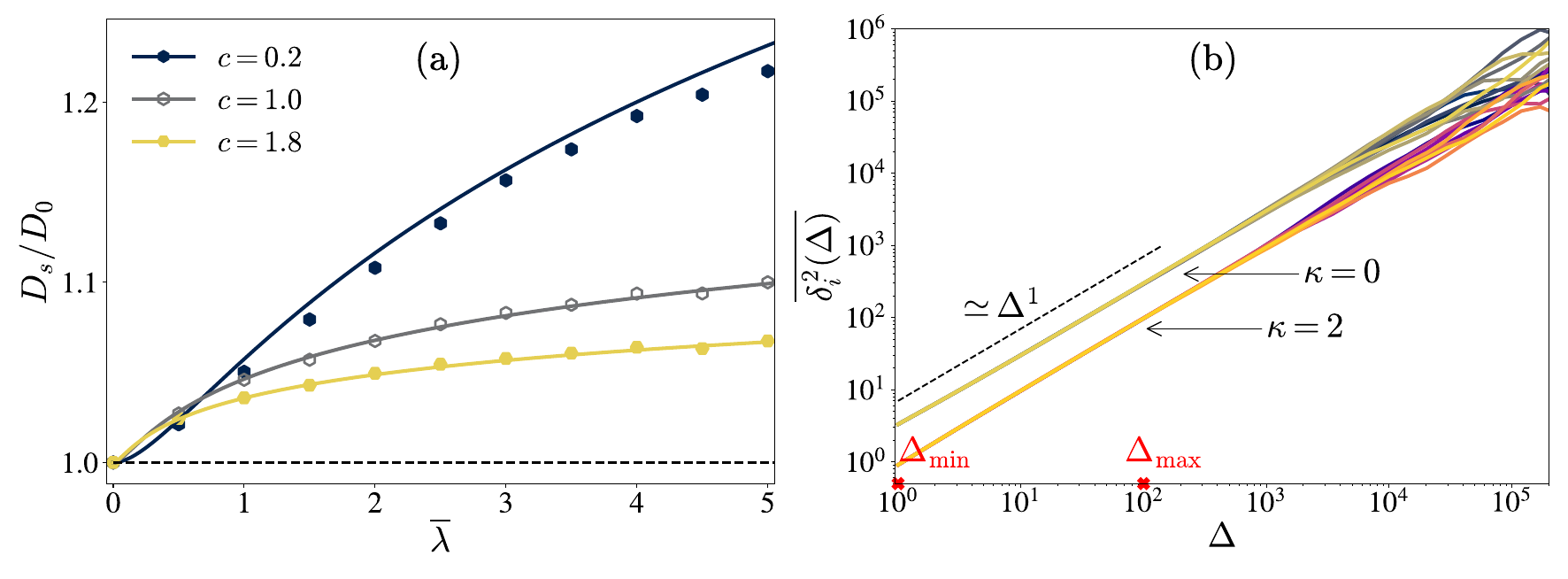}
   \caption{(a) Self-diffusion coefficient $D_\mathrm{s}/D_0$ as a function of 
   the dimensionless coupling parameter $\overline{\lambda}$. The curves are 
   computed with three different area fractions $c$, and with an oddness 
   parameter $\kappa=2>\kappa_c$, such that we are in the enhancement regime 
   and $D_\mathrm{s}>D_0$. 
   (b) Typical TAMSD $\overline{\delta^2_i(\Delta)}$ curves as a function of 
   lag-time $\Delta$ related to $i=1, \ldots, 10$ trajectories of a total 
   length of $T=4 \times 10^5$ real-time steps in the case of 
   $N=200$ interacting particles with density $c=0.1$ and $\kappa \in \{0, 2\}$. 
   The 
    values $\Delta_\mathrm{min}=1$ and $\Delta_\mathrm{max}=100$
    mark the interval of lag times $\Delta$ (in real-time units) used for 
    ensemble averaging the TAMSDs.}
    \label{fig:appendix}
\end{figure}

The conservative force $\bi{F}_i(t)$ exerted on particle $i$ and appearing 
in Eq. \eref{eq:velocity_SDE} is given by the 
sum of pairwise interaction forces $\bi{F}_i(t) = \sum_{j=0}^N \bi{f}_{ij}(t)$, 
where $\bi{f}_{ij}$ derives from a (normalised) Gaussian interaction potential 
$\mathcal{U}$, i.e.,
$ \bi{f}_{ij} = - \lambda\nabla \mathcal{U}(R_{ij})$ if $R_{ij} < \delta_c$ 
and $\bi{f}_{ij}=0$ if $R_{ij} > \delta_c$. Here, $R_{ij} = |\bi{X}_i(t) - 
\bi{X}_j(t)|$ 
is the inter-particle distance, and $\delta_c$ denotes a cut-off length scale 
that we use in order to truncate the interaction potential $\mathcal{U}(R_{ij})$ 
for reducing the computational time required by the Brownian dynamics 
simulations. In particular, if $\delta_c$ is sufficiently 
larger than the typical decay length of $\mathcal{U}(R_{ij})$, the error 
introduced 
by this truncation is negligible.
For the Gaussian interaction potential reported in 
Eq.~\eref{eq:Gaussian_interaction_potential},
we choose $\sigma=1.0$. Note that $\sigma$ is used as an effective 
particle radius, from where we deduce the effective concentration of particles 
$c = \pi \sigma^2\,  N / L^2$, where $L$ is the length of the square simulation box. 
The cut-off distance of the interaction force is chosen to be $\delta_c=4\sigma$. 

The dimensionless interaction scale $\overline{\lambda} = \lambda_\mathrm{tr}/
(2\pi \sigma^2\, T)$ is given in terms of the coupling $\lambda_\mathrm{tr}$ 
compared with the thermal energy $T$ and the length scale $\sigma$ of the 
Gaussian potential. For a comparison between theoretical predictions and 
simulation results, see 
Fig.~\ref{fig:appendix}(a), where different couplings $\overline{\lambda}$ are 
tested. The analytical predictions are expected to be valid in a regime where 
the coupling between the tracer and the host particle is sufficiently small. 
However, from Fig.~\ref{fig:appendix}(a) it can be seen that at high densities 
($c=1.0$ and $c=1.8$) the whole range of tested couplings produces very accurate 
results. At the low density $c=0.2$, instead, the mismatch between the 
analytical prediction and the 
numerical data increases upon increasing $\overline{\lambda}$. As a 
compromise between 
the accuracy of the analytical predictions and the magnitude of the effects shown 
in the figures of the main text (Section~\ref{sec:self_diffusion}), we opt 
for $\overline{\lambda}=1$ and compare the results with $\overline{\lambda}=4$ 
in Fig.~\ref{fig:Ds_vs_c} of the main text. Note that  $\overline{\lambda}=1$ 
is such that the maximum 
of the interaction energy $\overline{\lambda}\, \mathcal{U}(R_{ij})$ is equal 
to the thermal 
energy $T$. 

To solve the first-order stochastic differential equation \eref{eq:velocity_SDE} 
we use the standard Euler-Maruyama scheme \cite{kalz2022collisions}, where 
$\bi{X}_i(t_{n+1})$ and $\bi{V}_i(t_{n+1})$ are calculated from $\bi{X}_i(t_{n})$ 
and $\bi{V}_i(t_{n})$ and $t_{n+1}=t_n + \Delta t$ with $\Delta t = 10^{-3}$. 
The thermal noise is accounted for by $\sqrt{2\gamma_i T \Delta t}\, 
\mathcal{N}(0,\mathbf{1})$, where $\mathcal{N}(0,\mathbf{1})$ is a two-dimensional 
random vector drawn from a multivariate normal distribution with zero mean and 
covariance matrix given by the identity matrix $\mathbf{1}$. Note that the 
discretised stochastic equations of motion are interpreted according to the 
It\^o prescription, implying that the standard deviation of the noise is 
proportional to $\sqrt{\Delta t}$. 
To simulate Eq.~\eref{eq:velocity_SDE} we use a square box of length $L$ with 
periodic boundary conditions, where the box length is determined so 
as to result into the desired density of particles $c$, i.e., $L= \sqrt{\pi 
\sigma^2 N/c }$. 
As the interaction force does not diverge for $R_{ij} \rightarrow 0$ and 
particle 
overlaps are possible, we initialise 
the position of the $N=200$ particles according to a uniform distribution over 
the finite box. After an initial equilibration period of $n_\mathrm{eq} = 
10^7$ time steps, we start recording the stochastic trajectory for a total 
duration of $n_\mathrm{tot} = 4 \times 10^8$ time steps, which corresponds to a 
trajectory length of $T = 4 \times 10^{5}$ in real-time units. For an efficient 
computation, we used a neighbour-list implementation for the evaluation of the 
interaction forces, with a buffer radius $\delta_\mathrm{buff}$ which has been 
optimised in order to minimise the computational time. Over the broad range of 
densities simulated, a buffer-radius of $\delta_\mathrm{buff} \approx 2 \delta_c$ 
turned out to be the most efficient.

\subsection{MSD evaluation}

In order to evaluate the diffusion coefficient of the tracer particle 
we calculate the time-averaged MSD (TAMSD) \cite{he2008random} for each 
(independent) trajectory $i$ 
according to 
\begin{eqnarray}
    &\overline{\delta^2_i(\Delta, T)} = \frac{1}{T-\Delta}  \int_{0}^{T -\Delta}
    \mathrm{d}t\ 
    |\bi{X}_{0}^{\{i\}}(t + \Delta) - \bi{X}_{0}^{\{i\}}(t)|^2,
\end{eqnarray}
where $\bi{X}_{0}^{\{i\}}(t)$ is the the position of the tracer particle at time 
$t$ in trajectory $i$, $T$ is the trajectory length and $\Delta$ is the lag time. 
As the system under consideration is ergodic, we can 
ensemble-average over the $i=1, \ldots, i_\mathrm{max}$ independent trajectories 
to obtain the estimate for the MSD 
\cite{burov2011single}, which is formally defined as
\begin{eqnarray}
    \left\langle |\mathbf{X}_0(t) - \mathbf{X}_0(0)|^2\right\rangle &= 
    \lim_{T\to\infty} \left\langle \overline{\delta^2_i(\Delta=t, T)}\right\rangle 
    \nonumber \\
    & =  \lim_{T\to\infty} \frac{1}{i_\mathrm{max}} \sum_{i=1}^{i_\mathrm{max}} 
    \overline{\delta^2_i(t=\Delta, T)}.
\end{eqnarray}
By taking $T$ large enough one can 
assume that the initial conditions play no role in the evaluation 
of the long-time MSD. Hence, for the sake of simplicity, we impose 
$\mathbf{X}_0(t_0)=\mathbf{0}$. We observe from Fig.~\ref{fig:appendix}(b) 
that the most reliable $\Delta$-range from which to extract the MSD is 
$1=\Delta_\mathrm{min} \leq \Delta \leq \Delta_\mathrm{max}=100$. The MSD is 
then used to deduce the self-diffusion coefficient $D_\mathrm{s}$ by fitting a 
linear time-dependence $\langle |\mathbf{X}_0(t)|^2\rangle = 4\, D_\mathrm{s}\, 
t$, where we take $N_\Delta=40$ logarithmically equidistant lag-times to fit 
the MSD and ensemble average over $i_\mathrm{max}=10$ independent trajectories. 

\section*{References}


\end{document}